\documentclass[aip,jcp,reprint]{revtex4-2}

\pdfoutput=1
\usepackage[utf8]{inputenc}
\usepackage[english]{babel}
\usepackage[T1]{fontenc}
\usepackage{csquotes}
\usepackage{physics}
\usepackage{amsmath,amsfonts}
\usepackage{cancel}

\usepackage[dvipsnames]{xcolor}
\usepackage[table]{xcolor} 

\usepackage[colorlinks=true,linkcolor=blue,citecolor=blue,urlcolor=blue]{hyperref}
\usepackage[capitalise]{cleveref}

\usepackage{listings}

\usepackage{multirow}
\usepackage{graphicx}
\graphicspath{{./}}

\usepackage{bm} 

\usepackage[most]{tcolorbox}
\tcbset{
  jcpbox/.style={
    colback=gray!5,       
    colframe=black!40,    
    boxrule=0.5pt,
    arc=2pt,
    left=6pt,right=6pt,top=4pt,bottom=4pt
  }
}
\tcbuselibrary{theorems}
\newtcbtheorem{mybox}{Box}{
  colback=gray!5,
  colframe=gray!60!black,
  fonttitle=\bfseries
}{box}

\crefname{tcb@cnt@mybox}{Box}{Boxes}

\begin{document}

\title{
Variance reduction for forces and pressure in variational Monte Carlo
}

\author{David Linteau}
\email{david.linteau@epfl.ch}
\affiliation{Institute of Physics, \'{E}cole Polytechnique F\'{e}d\'{e}rale de Lausanne (EPFL), CH-1015 Lausanne, Switzerland}
\affiliation{Center for Quantum Science and Engineering, \'{E}cole Polytechnique F\'{e}d\'{e}rale de Lausanne (EPFL), CH-1015 Lausanne, Switzerland}

\author{Saverio Moroni}
\affiliation{CNR-IOM DEMOCRITOS, Istituto Officina dei Materiali and SISSA Scuola Internazionale Superiore di Studi Avanzati, Via Bonomea 265, I-34136 Trieste, Italy}

\author{Giuseppe Carleo}
\affiliation{Institute of Physics, \'{E}cole Polytechnique F\'{e}d\'{e}rale de Lausanne (EPFL), CH-1015 Lausanne, Switzerland}
\affiliation{Center for Quantum Science and Engineering, \'{E}cole Polytechnique F\'{e}d\'{e}rale de Lausanne (EPFL), CH-1015 Lausanne, Switzerland}

\author{Markus Holzmann}
\affiliation{Univ. Grenoble Alpes, CNRS, LPMMC, 38000 Grenoble, France}

\begin{abstract}

We present simple and practical strategies to reduce the variance of Monte Carlo estimators.
Our focus is on variational Monte Carlo calculations of atomic forces and pressure in electronic systems, although we show that the underlying ideas apply more broadly to other observables, like pair-correlation and angular-distribution functions, and other methods, including molecular dynamics.
For Pulay-type contributions, we show that a minor modification based on the Metropolis acceptance ratio softens the power-law divergence of the variance to a logarithmic one, thus reducing the tails of the distribution.
Proposal averaging can be added to remove completely this nodal divergence, or, alternatively, inexpensive regularizations can be applied to suppress remaining outliers at the price of a controlled small bias.
For Hellmann-Feynman forces, we derive compact variance-reduced estimators for periodic systems that are straightforward to implement in standard Monte Carlo codes. 
The approach is illustrated for high-pressure metallic hydrogen with more than a hundred atoms described by neural quantum states, including an application to molecular dynamics driven by the improved forces.

\end{abstract}

\maketitle

\section{Introduction}

Accurate evaluation of atomic forces and the stress tensor is central to first-principles studies of quantum many-body systems.
Forces determine equilibrium geometries, vibrational properties, and the atomic-scale dynamics in molecules and condensed matter systems \cite{1939_feynman_forces,1969_pulay_force}.
Likewise, the stress (or pressure) underpins equations of state, elastic responses, and structural phase transitions \cite{1983_nielsen_martin_stress,1985_nielsen_martin_stress_and_force}.
Among first-principles approaches, Variational Monte Carlo (VMC) offers a simple yet systematically improvable framework for describing strongly correlated electrons and quantum nuclei, especially with neural quantum state (NQS) wave functions that now routinely reach the accuracy of previous diffusion Monte Carlo (DMC) calculations \cite{2020_spencer_ferminet,2020_hermann_paulinet,2023_cassella_ferminet_electron_gas,2023_wilson_nqs_electron_gas,2024_pescia_mpnn_electrongas,2024_kim_ultra_cold_fermi_gases,2024_smith_2deg,2025_linteau_helium4_2d,2025_linteau_atomic_hydrogen,2026_smith_moire_wigner_crystal}.

Within VMC, it is possible to obtain force fields $\mathbf{F}(\mathbf{R}) = -\nabla_\mathbf{R} E(\mathbf{R})$
which are consistent with the variational Born-Oppenheimer energy surface $E(\mathbf{R})=\langle \Psi(\mathbf{R}) | H(\mathbf{R}) | \Psi(\mathbf{R}) \rangle / \langle \Psi(\mathbf{R}) | \Psi(\mathbf{R}) \rangle$.
Here, $|\Psi(\mathbf{R}) \rangle$ represents the variational approximation to the ground state of the electronic Hamiltonian $H(\mathbf{R})$, both of which depend parametrically on the nuclear positions $\mathbf{R}$. 
The total derivative with respect to nuclear position can then readily be calculated, leading to consistent conservative force fields (up to unavoidable stochastic noise). 
Foundation model wave functions \cite{2024_scherbela_transferable_fermionic_neural_wf,2024_gao_neural_pfaffian,2025_rende_foundation_neural_network_quantum_states,2025_foster_orbformer,2026_szabo_fnqs_chemical_processes} are particularly attractive because they deliver well-defined and consistent forces across many nuclear configurations at far lower cost than re-optimizing from scratch at each geometry.
While it is possible to obtain consistent forces within DMC \cite{2022_moroni_forces_by_regularization,2024_slootman_qmc_forces_ethanol}, it involves averaging (and extrapolating) over the imaginary time history, which is avoided in a pure VMC framework.

However, directly differentiating the VMC energy expectation value typically leads to large statistical fluctuations.
These arise both from the singular behavior of estimators at electron-nucleus coalescence points, and from the electronic wave function approaching the nodes.

A variety of approaches have already been proposed to address these challenges \cite{2008_kalos_whitlock_monte_carlo_methods_book}.
Early work focused on coordinate transformations and correlated sampling, notably the space-warp coordinate transformation \cite{1989_umrigar_pes_swct,2000_filippi_swap_warp_forces,2024_qian_space_warp_forces}.
A second line of work constructed renormalized estimators obeying the zero-variance/zero-bias principle \cite{1999_assaraf_caffarel_zero_var,2000_assaraf_caffarel_improved_forces,2003_assaraf_caffarel_zero_variance_force,2007_toulouse_zero_var_bias_pair_density}.
Finally, other methods proposed isolating and fitting the short-range singular contribution, possibly combined with antithetic sampling \cite{2005_chiesa_force_estimator}.

These methods have mostly focused on estimators for
the Hellmann-Feynman force $-\nabla_\mathbf{R} H(\mathbf{R})$, where
they significantly reduce the variance and improve the reliability of the results.
Methods to cure the variance problems occurring in the Pulay (or wave function) term
due to the parametric dependence of the approximate ground state wave function have been addressed differently.
They either modify the weight of the Monte Carlo sampling through a guiding wave function \cite{1988_ceperley_bernu_excited_state_guiding_wfc,2008_attaccalite_md_with_forces_from_qmc,2026_misery}, introduce a regularization leading to a small systematic bias \cite{2020_pathak_wagner_gradient_regularization},
or introduce a warp regularization that requires the associated Jacobian and derivatives of the local energy \cite{2022_moroni_forces_by_regularization}.
We also note that other lines of work address heavy-tailed QMC statistics through alternative sampling and tail-aware error analysis \cite{2008_trail_heavy_tails_in_qmc,2008_trial_alternative_sampling_for_vmc,2019_lopez_conduit_tail_regression_qmc}.

In this work, we focus on variance issues in both the Hellmann–Feynman and Pulay contributions, and propose practical estimators for forces and pressure in periodic systems, with straightforward extensions to open boundaries.
Building on earlier developments, especially by Assaraf and Caffarel \cite{1999_assaraf_caffarel_zero_var,2000_assaraf_caffarel_improved_forces,2003_assaraf_caffarel_zero_variance_force}, we derive expressions that add minimal (implementation) overhead in standard VMC codes and avoid introducing system-dependent tuning parameters or externally chosen numerical cutoffs for improved Hellmann-Feynman estimators.

For the Pulay terms, we revisit an acceptance trick proposed nearly half a century ago \cite{1977_ceperley_monte_carlo_fermions} and show that it can attenuate the variance spikes induced by the nodal surface, improving the variance scaling from polynomial to logarithmic with the distance to the nodes.
This acceptance trick provides a systematic and essentially cost-free route to reducing the variance of a wide class of estimators.
We then introduce a stronger proposal-averaged estimator that removes the nodal divergence completely. 
Alternatively, regularization schemes \cite{2020_pathak_wagner_gradient_regularization,2022_moroni_forces_by_regularization} can be used so that the residual bias remains negligible compared with the associated statistical uncertainty.

Although we focus on forces and pressure in the presentation and the numerical results, the methods described here apply more broadly to any parameter-dependent expectation values.
This includes, in particular, energy optimization, time-dependent variational Monte Carlo \cite{2012_Carleo,2017_Carleo,2020_Schmitt,2021_Verdel,2023_Sinibaldi,2024_Nys,2025_Gravina}, and the infinite-variance problems that arise generically in fermionic quantum Monte Carlo calculations \cite{2016_Shi,2025_Wan}.
We also discuss a related strategy that extends beyond derivative-based estimators to density-like observables sampled in Monte Carlo or molecular dynamics.

In the following, we first derive the general decomposition of energy derivatives into Hellmann–Feynman and Pulay contributions, and specialize the resulting expressions to atomic forces and pressure. 
We then introduce several practical variance-reduction strategies, including covariance forms for Pulay estimators, acceptance-weighted and proposal-averaged treatments of nodal divergences, and integrations-by-parts constructions for Hellmann-Feynman forces in both open and periodic systems.
Finally, we illustrate these ideas on high-pressure hydrogen described by a NQS, including applications to pressure, forces, molecular dynamics and pair correlations.
Most derivations and additional details, including near-node scaling, tail power-law exponents and moment-existence criterion, as well as a toy model example and a time-dependent variational Monte Carlo example, are provided in \cref{app:near-node-scaling} or in the \textit{supplementary material}.

\section{Variational Monte Carlo: naive estimators} \label{sec:default_vmc_estimators}

Consider a Hamiltonian $H(p)$ depending on a set of parameters $p$, and an explicit (not necessarily normalized) state $|\Psi(p)\rangle$ defined consistently with the chosen boundary conditions for all $p$ of interest.
Within VMC, we sample the electronic configurations $\mathbf{r} = (\mathbf{r}_1,\dots,\mathbf{r}_N)$ from the probability density $\pi(\mathbf{r};p) \equiv |\Psi(\mathbf{r};p)|^2 / (\int d\mathbf{r} |\Psi(\mathbf{r};p)|^2)$, where $\Psi(\mathbf{r};p) = \langle \mathbf{r} | \Psi(p) \rangle$ is the parameter dependent electronic wave function. 
For any observable $A$, the standard VMC estimator is the corresponding local quantity $A_L = \langle \mathbf{r}|A|\Psi(p)\rangle / \Psi(\mathbf{r};p)$. 
The variational energy is obtained by averaging local energy instances evaluated at configurations distributed according to the density $\pi(\mathbf{r};p)$, that is,
\begin{equation}
	E(p) = \langle H(p) \rangle = \mathbb{E}_{\mathbf{r} \sim \pi(\mathbf{r};p)} \left[ E_L(\mathbf{r};p) \right], 
\end{equation}
where the local energy is defined as follows
\begin{equation}
	E_L(\mathbf{r};p) = \frac{\langle \mathbf{r} | H(p) |\Psi(p)\rangle}{\Psi(\mathbf{r};p)}.
\end{equation}
We will often abbreviate configuration-space averages over $\pi(\mathbf{r};p)$ using the bra-ket (expectation value) notation
\begin{equation}
    \langle A \rangle \equiv \mathbb{E}_{\mathbf{r} \sim \pi(\mathbf{r};p)}[A_L(\mathbf{r};p)],
\end{equation}
and drop the explicit dependence on $p$.

In the case of nuclear forces and pressure, discussed below, 
we will consider the electronic Hamiltonian at fixed nuclear positions $\mathbf{R}$,
\begin{equation} \label{eq:electronic_hamiltonian}
H(\mathbf{R}) = K_e(\mathbf{r}) + V_{ee}(\mathbf{r}) + V_{en}(\mathbf{r}; \mathbf{R})+V_{nn}(\mathbf{R}),
\end{equation}
where $K_e(\mathbf{r}) = -\hbar^2/(2m) \sum_i \nabla_{\mathbf{r}_i}^2$ is the electronic kinetic energy operator.
Here $V_{ee}$, $V_{en}$, and $V_{nn}$ are the electron-electron, electron-ion, and ion-ion Coulomb interactions, respectively, for electrons of charge $-e$ and nuclei of charge $Z_I e$ (with $I$ being the nuclear index).
Under the Born-Oppenheimer approximation, the nuclear degrees of freedom enter the Hamiltonian $H(\mathbf{R})$ only parametrically through their coordinates $\mathbf{R} = (\mathbf{R}_1, \hdots, \mathbf{R}_N)$.

\subsection{General parameter optimization} \label{sec:general_parameter_optimization}

Given $H(p)$ and $|\Psi(p)\rangle$, as defined in the previous section, we introduce the corresponding energy surface
\begin{equation}
	E(p)=\frac{\langle \Psi(p) | H(p) |\Psi(p)\rangle}{\langle \Psi(p)|\Psi(p)\rangle},
\end{equation}
which, for each fixed $p$, is an upper bound to the exact ground state energy of $H(p)$.

We are not only interested in the energy $E(p)$ for given values of $p$, but also
in its derivative with respect to $p$. 
Assuming a sufficiently smooth parameter dependence of the
Hamiltonian and the variational wave function, we have

\begin{equation} \label{eq:dEdp}
    \frac{\partial E}{\partial p} = \underbrace{\frac{\langle \Psi | \partial_p H | \Psi \rangle} {\langle \Psi|\Psi\rangle}}_{\substack{\text{Hellmann-Feynman} \\ \text{(operator) term}}}
    +
    2 \mathrm{Re} \underbrace{\frac{\langle \Psi | H - E | \partial_p \Psi \rangle} {\langle \Psi|\Psi\rangle}}_{\substack{\text{Pulay} \\ \text{(wave function) term}}}.
\end{equation}
In the case where $|\Psi(p)\rangle$ is proportional to an exact eigenstate, so that $H(p)|\Psi(p)\rangle = E(p) |\Psi(p)\rangle$, the Pulay term clearly vanishes. 
More generally, it is sufficient that the energy functional $E(p,|\phi\rangle) = \langle \phi| H(p) |\phi \rangle/ \langle \phi |\phi \rangle$ be stationary, at fixed $p$, with respect to variations along the direction $|\partial_p \Psi \rangle$, i.e. within the subspace $|\phi\rangle = |\Psi(p) \rangle + \lambda |\partial_p \Psi \rangle$.
Although this stationarity condition may be satisfied at isolated parameter values $p_0$, enforcing it
along an entire continuous path in parameter space is much more restrictive.
As a result, both the Hellmann-Feynman and Pulay contributions should generally be retained 
to obtain the correct energy derivative.

For pure energy-minimization problems, the Hamiltonian is parameter independent, so the Hellmann-Feynman contribution vanishes and only the Pulay term remains,
\begin{equation}
	\frac{\partial E(p)}{\partial p}
	= 2 \mathrm{Re} \ \Big\langle [E_L(\mathbf{r}) - E(p)] \partial_p \log \Psi^*(\mathbf{r};p) \Big\rangle.
    \label{eq:pulay}
\end{equation}
Since this term can be written as a covariance (a form that we will leverage in \cref{sec:covariance}),
\begin{equation}
    \frac{\partial E(p)}{\partial p} = 2 \mathrm{Re} \ \mathrm{Cov}(E_L, \partial_p \log \Psi^*),
\end{equation}
its magnitude is bounded by the product of the corresponding variances via the Cauchy-Schwarz inequality,
\begin{equation}
    \left| \frac{\partial E(p)}{\partial p} \right| \le 2 \sqrt{ \mathrm{Var}(E_L) \mathrm{Var}(\partial_p \log \Psi^*)}.
\end{equation}
This makes explicit how the zero-variance principle provides an analytical guarantee: as the variational wave function approaches an exact eigenstate, $\mathrm{Var}(E_L) \to 0$, so the Pulay contribution is automatically suppressed, provided $\mathrm{Var}(\partial_p \log \Psi^*)$ remains finite.

However, higher moments of the Pulay estimator in general do not exist.
Indeed, the Pulay estimator behaves as $|\Psi|^{-2}$ approaching the nodes,
\begin{equation} \label{eq:pulay_nodal_divergence}
    \partial_p E \sim E_L \partial_p \log \Psi^* \sim \left( \frac{\nabla_\mathbf{r}^2 \Psi}{\Psi} \right) \left( \frac{\partial_p \Psi^*}{\Psi^*} \right) \sim |\Psi|^{-2},
\end{equation}
given that the Laplacian of the wave function, $\nabla_\mathbf{r}^2 \Psi(\mathbf{r};p)$, as well as $\partial_p \Psi^*(\mathbf{r};p)$
generally do not vanish at the nodes of $\Psi(\mathbf{r};p)$.
The mean remains finite because the sampling weight $\pi \propto |\Psi|^2$ suppresses this behavior at the nodes.
On the other hand, the second moment of the Pulay estimator, involving $|\Psi|^{-4}$, will cause the variance to diverge.
This divergence will generically show up in all Pulay terms occurring in parameter derivatives of general observables involving fermionic wave functions.
We will show below how an acceptance trick helps to control the derivative in practice.

So far, we have focused on pure energy minimization, for which the Hamiltonian is parameter independent and only the Pulay term survives. 
This changes for observables such as atomic forces and pressure, which are obtained by differentiating the energy with respect to parameters that enter the Hamiltonian explicitly, so that both Hellmann–Feynman and Pulay terms contribute.
We now turn first to forces and then to pressure.

\subsection{Nuclear forces}

Within the Born-Oppenheimer approximation, recall that the nuclear coordinates $\mathbf{R}$ enter the electronic Hamiltonian $H(\mathbf{R})$, defined in \cref{eq:electronic_hamiltonian}, only as parameters.
The Hellmann-Feynman force on nucleus $I$ then reads
\begin{equation}
    \mathbf{F}_I^\text{HF} = -\langle \nabla_{\mathbf{R}_I} H(\mathbf{R}) \rangle.
\end{equation}
Stochastic fluctuations will only come from the electron-nucleus term
\begin{align}
     - \langle \nabla_{\mathbf{R}_I} V_{en} \rangle &= -\sum_i \big\langle \nabla_{\mathbf{R}_I}v_{en}(\mathbf{R}_I - \mathbf{r}_i) \big\rangle
     \\
     &= \int d^3\mathbf{u} \ \mathbf{f}(\mathbf{u})\rho(\mathbf{R}_I-\mathbf{u})
     \label{eq:F_Ien_HF_v1},
\end{align}
where $\mathbf{f}(\mathbf{u})=-\nabla v_{en}(\mathbf{u})$ is the electron-nucleus force kernel and $\rho(\mathbf{u}) = \langle \sum_{i=1}^N \delta(\mathbf{u} - \mathbf{r}_i) \rangle$ is the electronic density. 

To keep the expressions compact, we introduce the electron-nucleus relative vector as $\mathbf{x}_{iI} \equiv \mathbf{r}_i - \mathbf{R}_I$, the associated distance $x_{iI} = |\mathbf{x}_{iI}|$, and the $\alpha$-component $x_{iI,\alpha} \equiv (\mathbf{x}_{iI})_\alpha$.
When no confusion can arise, we omit particle indices and simply write $\mathbf x$, with $x = |\mathbf{x}|$.
For isolated systems (open boundary conditions), $v_{en}(x) \sim 1/x$, while for extended systems (periodic boundary conditions), $v_{en}$ must be modified to account for the interaction with periodic images, which is typically handled using Ewald summation \cite{1921_ewald}.
In practice, this is achieved by splitting the interaction into short- and long-range parts,
\begin{equation} \label{eq:v_en_sr_lr_split}
    v_{en}(x)=v_{en}^{sr}(x)+v_{en}^{lr}(x),
\end{equation}
where $v_{en}^{sr}$ retains the $1/x$ behavior at small $x$, while the long-range component $v_{en}^{lr}$ is evaluated in Fourier space.

While the long-range contribution to the Hellmann-Feynman force is typically well behaved, the short-range part leads to a singular force kernel $\mathbf{f}(\mathbf{x}) \sim \mathbf{x}/x^3$ as $x \to 0$.
Since the electronic density remains finite at the nucleus, $\rho(\mathbf{R}_I - \mathbf{u}) \simeq \rho(\mathbf{R}_I) \ne 0$, the integral in \cref{eq:F_Ien_HF_v1} remains finite, and $\mathbf{F}^\mathrm{HF}_I$ is therefore an unbiased estimator of the Hellmann-Feynman force.
The variance, however, diverges because the second moment involves $\mathbf{f}^2(\mathbf{u}) \sim u^{-4}$, yielding
\begin{equation}
    \int d^3\mathbf{u} \ \mathbf{f}^2({\mathbf{u}}) \rho(\mathbf{R}_I - \mathbf{u})\simeq 4 \pi \rho(\mathbf{R}_I)\int_0^\epsilon u^{-2} du \to \infty,
\end{equation}
as $\epsilon \to 0$.
In this work, we will show how this divergence can be cured in simple ways.

The Pulay force has the general form given in \cref{eq:pulay}, and its variance diverges, as described in \cref{sec:general_parameter_optimization}, due to the fermionic nodes.
In practice, however, this divergence is usually not severe for forces given that $\partial_{R_{I\alpha}}$ acts locally (as compared to the pressure as we will discuss next), and as long as the fluctuations of $\Delta E \equiv E_L(\mathbf{r}) - E(p)$ away from 0 are not large (which is typically the case when the variational wave function is close to the ground state).
This will be shown in the Results section.

\subsection{Pressure} \label{sec:pressure}

As a second key observable, we consider the thermodynamic pressure, $P = -\partial E / \partial \Omega$, where $\Omega = L_x L_y L_z$ is the volume of an orthorhombic simulation cell of size $\mathbf{L} = (L_x, L_y, L_z)$.
(This choice is made for clarity; the  derivation generalizes straightforwardly to a non-orthorhombic cell by replacing $\mathbf{L}$ with the cell matrix $\mathbf{h}$ and $\Omega = \mathrm{det}\mathbf{h}$.)
An isotropic volume change is parameterized by a scale factor $s$ via $L_\alpha \to s L_\alpha$, for all $\alpha \in \{x,y,z\}$, so that $\Omega \to s^3 \Omega$ and
\begin{equation}
    P = - \frac{\partial E}{\partial \Omega} = - \left. \frac{1}{3\Omega} \frac{dE(s)}{ds} \right|_{s=1},
\end{equation}
where $E(s) \equiv E(\mathbf{R}(s), \mathbf{L}(s))$.

Although the box lengths can be viewed as parameters, this derivative cannot be taken naively using \cref{eq:dEdp}, because changing $\mathbf{L}$ modifies the boundary conditions of the wave function ($\Psi(\mathbf{r}_i+\mathbf{L}; \mathbf{L})= \Psi(\mathbf{r}_i; \mathbf{L})$ under periodic boundary conditions), which violates an implicit assumption behind \cref{eq:dEdp}.
To proceed, we therefore introduce scaled coordinates $\widetilde{\mathbf{r}}_i = \mathbf{r}_i \mathbf{L}^{-1}$ and $\widetilde{\mathbf{R}}_I = \mathbf{R}_I \mathbf{L}^{-1}$, so that the wave function of the scaled box reads
\begin{equation}
    \widetilde{\Psi}(\widetilde{\mathbf{r}}; \widetilde{\mathbf{R}},\mathbf{L}) = \Omega^{N/2} \Psi(\widetilde{\mathbf{r}}\mathbf{L}; \widetilde{\mathbf{R}}\mathbf{L},\mathbf{L}),
\end{equation}
with $\Omega^{N/2}$ being the square root of the Jacobian of the transformation $d\mathbf{r}_i = \Omega d\tilde{\mathbf{r}}_i$ (for all $1 \le i \le N$) needed so that $|\widetilde{\Psi}|^2 d\tilde{\mathbf{r}}$ represents the same probability measure as $|\Psi|^2 d\mathbf{r}$.
The associated scaled Hamiltonian is
\begin{align}
    \widetilde{H}(\widetilde{\mathbf{R}},\mathbf{L}) &=
    \widetilde{K}_e(\widetilde{\mathbf{r}}\mathbf{L})
    +V_{ee}(\widetilde{\mathbf{r}}\mathbf{L}) \nonumber\\
    &\quad + V_{en}(\widetilde{\mathbf{r}}\mathbf{L}; \widetilde{\mathbf{R}} \mathbf{L})
    +V_{nn}(\widetilde{\mathbf{R}}\mathbf{L}),
\end{align}
where $\widetilde{K}_e(\widetilde{\mathbf{r}}) = -\hbar^2/(2m) \sum_{i,\alpha} L_{\alpha}^{-2} \partial_{\widetilde{r}_{i\alpha}}^2$. 
By construction, $\widetilde{\Psi}$ satisfies the periodic boundary conditions on the unit box $[0,1)^3$, independently of $\mathbf{L}$, so partial derivatives with respect to $\mathbf{L}$ (or $s$) are therefore well defined.
We can thus apply the general parameter-derivative formula given in \cref{eq:dEdp} to $\widetilde{H}$ and $\widetilde{\Psi}$, and then re-express the final results in terms of the original $H$ and $\Psi$.

Defining the shorthands $\widetilde{\Psi}(s) \equiv \widetilde{\Psi}(\, \cdot \, ; \widetilde{\mathbf{R}},\mathbf{L}(s))$
and $\widetilde{H}(s)\equiv \widetilde{H}(\widetilde{\mathbf{R}},\mathbf{L}(s))$, the variational
energy reads
\begin{equation}
    E(s) = \frac{\langle \widetilde{\Psi}(s) |\widetilde{H}(s) |\widetilde{\Psi}(s) \rangle}
    {\langle \widetilde{\Psi}(s) |\widetilde{\Psi}(s) \rangle},
\end{equation}
and its derivative decomposes into the usual Hellmann-Feynman (Virial) and wave function (Pulay) contributions,
\begin{equation}
    \frac{dE}{ds} =\big\langle \partial_s \widetilde{H}(s) \big\rangle_{\widetilde{\pi}} + 2 \mathrm{Re} \ \Big\langle (\widetilde{E}_L-E) \partial_s \log\widetilde{\Psi}^*(s) \Big\rangle_{\widetilde{\pi}},
\end{equation}
where $\widetilde{\pi}\propto|\widetilde{\Psi}(s)|^2$ and
$\widetilde{E}_L=(\widetilde{H}(s) \widetilde{\Psi}(s))/\widetilde{\Psi}(s)$.
The pressure splits accordingly, $P=P^V+P^{wf}$.
For pure Coulomb interactions, the explicit derivative of the 
scaled Hamiltonian, $\partial_s \widetilde{H}(s)$, 
yields the usual Virial estimator 
\cite{1985_nielsen_martin_stress_and_force}
\begin{equation}
    P^V=\frac{1}{3\Omega}\bigl(2K_e+V\bigr),
    \label{eq:Pvirial}
\end{equation}
with $V=V_{ee}+V_{en}+V_{nn}$.
In contrast to the nuclear forces above, the standard (unbiased) estimator for $P^V$ already has finite variance.

Rewriting the wave function contribution in terms of the original $\Psi$ gives
\begin{equation} \label{eq:pulay_pressure}
    P^{wf} = -\frac{2}{3\Omega} \mathrm{Re} \ \Big\langle \Delta E(\mathbf{r}; \mathbf{R},\mathbf{L}) \mathcal{D} \log\Psi^* (\mathbf{r}; \mathbf{R}, \mathbf{L}) \Big\rangle,
\end{equation}
where $\Delta E(\mathbf{r}; \mathbf{R},\mathbf{L})\equiv E_L(\mathbf{r}; \mathbf{R},\mathbf{L}) - E(\mathbf{R},\mathbf{L})$, and we conveniently introduced the dilation operator
\begin{equation} \label{eq:dilation_derivative}
    \mathcal{D} \equiv \left.\frac{d}{ds}\right|_{s=1}
    =\mathbf{r} \cdot \nabla_{\mathbf{r}}
    +\mathbf{R} \cdot \nabla_{\mathbf{R}}
    +\mathbf{L} \cdot \nabla_{\mathbf{L}},
\end{equation}
which, as the name suggests, is a generator of isotropic scaling (preserving the relative geometry within the cell).
Upon writing the scaled wave function explicitly along the path $\mathbf{L}(s)=s\mathbf{L}$,
\begin{equation}
    \widetilde{\Psi}(\tilde{\mathbf{r}}; \tilde{\mathbf{R}},\mathbf{L}(s)) = \Omega(\mathbf{L}(s))^{N/2} \Psi(\tilde{\mathbf{r}}\mathbf{L}(s); \tilde{\mathbf{R}}\mathbf{L}(s),\mathbf{L}(s)),
\end{equation}
one finds $\Omega(\mathbf{L}(s)) = s^3\Omega(\mathbf{L})$ and therefore
\begin{equation} \label{eq:dlog_tilde_psi_to_dlogpsi}
    \mathcal{D} \log\widetilde{\Psi}^* = \mathcal{D} \log\Psi^* + \frac{3N}{2}.
\end{equation}
We note that this construction can be extended beyond the pressure to the off-diagonal elements of the stress tensor.

\section{Variance reduction methods}

Above, we have seen that for forces, both the Hellmann-Feynman and Pulay terms have divergent second moments, while for pressure, the Pulay term is problematic.
As a result, the error of a Monte Carlo simulation with finite statistics cannot be estimated by standard methods based on the central limit theorem \cite{2017_Delyon}.
In the following, we present simple and practical methods for addressing these variance problems, still focusing on force and pressure calculations.
For detailed derivations and more sophisticated improvements, we refer to \cref{app:near-node-scaling} or the \textit{supplementary material}.

\subsection{Covariance} \label{sec:covariance}

Consider the standard Pulay estimator given in \cref{eq:pulay}, which, as hinted previously in \cref{sec:general_parameter_optimization}, can be recast in the equivalent covariance form

\begin{equation} \label{eq:pulay_covariance}
	\frac{\partial E}{\partial p} = 2 \mathrm{Re} \ \Big\langle \Delta E [ \partial_p \log \Psi^* - \langle\partial_p \log \Psi^*\rangle ] \Big\rangle.
\end{equation}
Crucially, the subtraction of $\langle\partial_p \log \Psi^*\rangle$ provides a simple but powerful variance-reduction trick, as it leaves the mean unchanged (since $\langle \Delta E \rangle = 0$) while typically reducing fluctuations substantially \cite{2005_sorella_wfc_optimization_in_vmc,2005_umrigar_filippi_energy_and_var_optimization}.

In the case of pressure, for example, where $\partial_p$ is replaced by the dilation derivative $\mathcal{D}$ introduced in \cref{eq:dilation_derivative}, this covariance formulation is especially valuable because it cancels the constant offset $3N/2 \sim \mathcal{O}(N)$ appearing in \cref{eq:dlog_tilde_psi_to_dlogpsi}, preventing this extensive constant from inflating the variance despite not contributing to the mean.

\subsection{Acceptance trick} \label{sec:acceptance_trick}

The Metropolis acceptance-ratio trick discussed in this section was already mentioned in the pioneering work of Ceperley, Chester and Kalos \cite{1977_ceperley_monte_carlo_fermions}.
To our knowledge, despite being introduced nearly half a century ago, its variance-reduction behavior for estimators suffering from nodal divergences has not been analyzed in detail.
We therefore take the opportunity to revisit it here, formalize its near-node behavior by showing how it reduces the variance divergence
of the naive Pulay type estimators from power-law to logarithmic (with respect to the distance to the nodal surface), and extend it to a regularized form with controlled bias.

The method exploits additional information available in the Metropolis-Hastings accept/reject step of the Markov chain.
Consider a proposed move from an initial configuration $\mathbf{r}_i$ to a proposed configuration $\mathbf{r}_f$.
The move is then accepted with probability 
\begin{equation}
    a \equiv a(\mathbf{r}_i \to \mathbf{r}_f) = \mathrm{min}\left(1, \frac{\pi(\mathbf{r}_f) p(\mathbf{r}_i | \mathbf{r}_f)}{\pi(\mathbf{r}_i) p(\mathbf{r}_f | \mathbf{r}_i)} \right),
\end{equation}
where $\pi$ is the target distribution, taken to be the Born distribution $\pi(\mathbf{r})=|\Psi(\mathbf{r})|^2$ in VMC, and $p(\mathbf{x}|\mathbf{y})$ is the proposal distribution to sample $\mathbf{x}$ given $\mathbf{y}$.
Instead of evaluating an observable $O(\mathbf{r})$ only at the configuration obtained after the accept/reject step, one can use the conditional estimator \cite{1977_ceperley_monte_carlo_fermions}
\begin{equation} \label{eq:acceptance_trick_estimator}
    \widetilde{O}(\mathbf{r}_i,\mathbf{r}_f) = a O(\mathbf{r}_f) + (1 - a) O(\mathbf{r}_i),
\end{equation}
which is defined independently of whether the proposal is ultimately accepted.
In other words, the acceptance probability is used to form a weighted average of the estimator evaluated at the initial and proposed configurations.

This construction is a simple instance of the Rao-Blackwell theorem \cite{1945_rao_information_and_accuracy,1947_blakwell_conditional_expectation,1996_casella_rao_blackwellisation_of_sampling,2008_kalos_whitlock_monte_carlo_methods_book}: rather than using the post-acceptance value $O(\mathbf{r}_{i+1})$, one replaces it by its conditional expectation given $(\mathbf{r}_i, \mathbf{r}_f)$.
Once the Markov chain has thermalized, this estimator remains unbiased,
\begin{align} \label{eq:tilde_O_unbiased}
    \mathbb{E}[\widetilde{O}(\mathbf{r}_i, \mathbf{r}_f)] 
    &= \mathbb{E}_{\mathbf{r}_i \sim \pi} \mathbb{E}_{\mathbf{r}_f \sim p(\mathbf{r}_f|\mathbf{r}_i)} \mathbb{E}_{A \sim \mathrm{Bern}}[O(\mathbf{r}_{i+1}|\mathbf{r}_i,\mathbf{r}_f)] \nonumber \\
    &= \mathbb{E}[O(\mathbf{r}_{i+1})] = \langle O \rangle,
\end{align}
where $A \in \{0,1\}$ is the Metropolis-Hastings acceptance indicator with $A \sim \mathrm{Bernoulli}(a(\mathbf{r}_i \to \mathbf{r}_f))$, ``Bern'' being a shorthand for a Bernoulli distribution, and $\mathbf{r}_{i+1} = A \mathbf{r}_f + (1-A) \mathbf{r}_i$.
Moreover, the Rao-Blackwell theorem guarantees that this replacement can only reduce the variance, or leave it unchanged (as derived explicitly in \cref{app:rao-blackwell_interpretation}), that is,
\begin{equation} \label{eq:rao_blackwell_O_tilde_vs_O}
    \mathrm{Var}( \widetilde{O}(\mathbf{r}_i,\mathbf{r}_f) ) \le \mathrm{Var}(O(\mathbf{r}_{i+1})).
\end{equation}

The acceptance trick can therefore be applied on top of essentially any estimator $O(\mathbf{r})$ as a cheap post-processing improvement to reduce numerical instabilities.
A stronger conditioning step is obtained by averaging over proposals $\mathbf{r}_f$ with the initial configuration $\mathbf{r}_i$ held fixed.
The resulting estimator,
\begin{equation} \label{eq:def_proposal_averaged_estimator}
    \overline{O}(\mathbf{r}_i) = \mathbb{E}[\widetilde{O}(\mathbf{r}_i, \mathbf{r}_f) | \mathbf{r}_i] = \int d^3 \mathbf{r}_f p(\mathbf{r}_f | \mathbf{r}_i) \widetilde{O}(\mathbf{r}_i, \mathbf{r}_f),
\end{equation}
remains unbiased and satisfies $\mathrm{Var}(\overline{O}) < \mathrm{Var}(\widetilde{O})$, as detailed in \cref{app:proposal_averaged_estimator}.
For Pulay-type observables, we show in this appendix that $\overline{O}(\mathbf{r}_i)$ removes the residual logarithmic nodal divergence of $\widetilde{O}(\mathbf{r}_i, \mathbf{r}_f)$ entirely.
In practice, $\overline{O}$ is approximated with $K$ proposals, leading to the estimator
\begin{equation} \label{eq:def_K_proposal_averaged_estimator}
    \overline{O}_K = \frac{1}{K} \sum_{k=1}^K \widetilde{O}(\mathbf{r}_i, \mathbf{r}_f^{(k)}),
\end{equation}
which interpolates between the acceptance estimator $(K=1)$ and full proposal averaging $(K \to \infty)$. 
Since $K$ may depend on $\mathbf{r}_i$, it can be chosen so that $K(\mathbf{r}_i) \to \infty$ as $\mathbf{r}_i$ approaches the nodal surface, eliminating completely the nodal divergence of Pulay-type observables.

In the following, we focus on the impact of the acceptance trick ($K=1$) on Pulay terms.
Our discussion concentrates on variance problems induced by nodal singularities, which are the most common source of large fluctuations.
We will always assume that $\partial_p \log \Psi$ has finite variance, and exclude too strong explicit singularities (which would require separate considerations,
analogous to those used for Hellmann-Feynman estimators discussed below).

\subsection{Regularization of Pulay terms}

We now build on top of the previous ``Acceptance trick'' Section, and focus on the general Pulay estimator given in \cref{eq:pulay}, of the form $O(\mathbf{r}) \sim (E_L(\mathbf{r}) - E) \partial_p \log \Psi^*(\mathbf{r})$, and analyze its behavior close to the nodes, where $\Psi(\mathbf{r}) \equiv \Psi(\mathbf{r};p) = 0$ (dropping the dependence on $p$ for this discussion).
In the vicinity of a node, we introduce local coordinates $\mathbf{r} = (\xi, \mathbf{r}_{\|})$, where $\xi$ is the normal coordinate, i.e. the signed distance to the nodal surface, and $\mathbf{r}_{\|}$ denotes the tangential coordinates.
We can then expand the wave function along the normal coordinate $\xi$ as $\Psi(\xi, \mathbf{r}_{\|}) = \partial_\xi \Psi(\xi, \mathbf{r}_{\|})|_{\xi=0} \ \xi + \mathcal{O}(\xi^2)$, so that the wave function vanishes linearly to leading order,
\begin{equation}
    \Psi(\xi, \mathbf{r}_{\|}) \approx \Phi(\mathbf{r}_{\|}) \xi,
\end{equation}
with $\Phi(\mathbf{r}_{\|}) \equiv \partial_\xi \Psi(\xi, \mathbf{r}_{\|})|_{\xi=0} \ne 0$, so that $|\Psi|^2 \sim \xi^2$.
Generically, both $H \Psi$ and $\partial_p \Psi$ approach non-zero constants as $\xi \to 0$. 
It follows that $E_L = H \Psi/\Psi \sim \xi^{-1}$ and $\partial_p \log \Psi^* = \partial_p \Psi^*/\Psi^* \sim \xi^{-1}$, which in turn implies that the Pulay estimator behaves as $(E_L - E) \partial_p \log \Psi^* \sim \xi^{-2}$.
Because the sampling weight vanishes as $|\Psi|^2 \sim \xi^2$, the mean remains finite.
The variance, however, involves the second moment $\xi^{-4}$ of the estimator, which leads to a divergent integral $\int \xi^2 \xi^{-4} d\xi$ and the naive power-law divergence scaling of $\mathcal{O}(1/\xi_c)$, where $\xi_c$ is a short-distance length scale introduced only to display the divergence of unregularized estimators.

A simple regularization scheme is to exclude a small region of width $\epsilon$ around the nodes by multiplying the integrand by a Heaviside step function $\theta_\epsilon(\xi) \equiv \theta(|\xi| - \epsilon)$, that vanishes for $|\xi| < \epsilon$.
This makes the variance finite, of order $\mathcal{O}(1/\epsilon)$, but introduces a bias, of order $\mathcal{O}(\epsilon)$, because the discarded region obviously contributes to the mean.
A clever choice of a smoother regularizing function \cite{2020_pathak_wagner_gradient_regularization} can actually reduce the bias to order $\epsilon^3$.

Let us instead use the acceptance trick together with a simple step-function regularization.
(While we apply the regularization to the acceptance trick $\widetilde{O} \equiv \overline{O}_{K=1}$, the same procedure holds for $\overline{O}_K$ for any finite $K$, whose value can vary as a function of $\mathbf{r}_i$.)
Rather than applying the cutoff symmetrically, we apply it only to the current configuration $\mathbf{r}_i$ inside the conditional estimator, that is, 
\begin{equation} \label{eq:hard_cutoff_regularized_pulay_pressure}
    \widetilde{O}_\epsilon(\mathbf{r}_i,\mathbf{r}_f)
    = \theta_\epsilon(\xi_i) \Big[ a O(\mathbf{r}_f) + (1 - a) O(\mathbf{r}_i) \Big],
\end{equation}
where $\xi_i$ is the orthogonal distance to the nodal surface of $\mathbf{r}_i$.
Inserting the near-node behavior of the wave function ($\Psi \sim \xi$) into this expression shows that the bias scales as $\mathcal{O}(\epsilon^2)$, while the variance grows only logarithmically, $\mathrm{Var}(\widetilde{O}_\epsilon) \sim \log (1/\epsilon)$.
This logarithmic behavior is a direct consequence of the acceptance trick, and should be contrasted with the naive estimator, whose variance scales as $\mathcal{O}(1/\epsilon)$.
Due to the very mild logarithmic divergence of the variance, one can choose $\epsilon$ small enough that the $\mathcal{O}(\epsilon^2)$ bias is negligible compared with statistical uncertainties.
Alternatively, one may eliminate the residual bias systematically by performing an explicit $\epsilon \to 0$ extrapolation, as proposed by Pathak and Wagner \cite{2020_pathak_wagner_gradient_regularization}.

In \cref{app:near-node-scaling} and in the \textit{supplementary material}, we analyze these asymptotic scalings in detail for the acceptance estimator $\widetilde{O}$, the proposal-averaged estimator $\overline{O}$ (and its finite-$K$ approximation), and the three regularized acceptance-weighted estimators introduced there: (1) the ``one-point'' hard cutoff $\widetilde{O}_\epsilon^{(1)}$, corresponding to \cref{eq:hard_cutoff_regularized_pulay_pressure}, (2) the ``two-point'' hard cutoff $\widetilde{O}_\epsilon^{(2)}$ (generalizing the cutoff $\theta_\epsilon(\xi_i)$ used in \cref{eq:hard_cutoff_regularized_pulay_pressure} to also include $\xi_f$), (3) and the smooth-cutoff estimator $\widehat{O}_\epsilon$.
In particular, we show that the two hard-cutoff forms retain the favorable logarithmic variance while introducing a $\mathcal{O}(\epsilon^2)$ bias, and that this bias can be systematically improved by replacing such ``hard-cutoffs'' with a smooth polynomial regularization function $\chi$.
If one then imposes the endpoint and smoothness conditions $\chi(0)=0, \chi'(0)=0, \chi(1)=1$ and $\chi'(1)=0$, a polynomial cutoff of degree $m+3$ is sufficient to cancel the first $m$ bias moments and as a result obtain a residual bias of order $\epsilon^{m+2}$ for $m \ge 1$ (see also Ref.~\cite{2020_pathak_wagner_gradient_regularization}).

\subsection{Integration by parts (for Hellmann-Feynman forces)} \label{sec:ibp_for_hf_forces}

In contrast to the Pulay force, the naive estimator of the Hellmann-Feynman (HF) force does not satisfy a zero-variance principle, meaning that the operator still fluctuates even when the wave function sampled coincides with an exact energy eigenstate.
Several variance-reduction strategies have been proposed \cite{2000_assaraf_caffarel_improved_forces,2000_filippi_swap_warp_forces,2003_assaraf_caffarel_zero_variance_force,2005_chiesa_force_estimator}.
In this section, we present a simple two-step approach applicable both to isolated (OBC) and extended (PBC) systems that relies on integrating by parts (IBP).
The first step partially removes the coalescence singularity to leading order, while the second cancels it exactly by exploiting a ``gradient-to-Laplacian'' identity. 

This key identity, rapidly derived in Sec. II of the \textit{supplementary material}, states that for any functions $v(x)$ and $f(x) = x v(x)$, provided $x \ne 0$,
\begin{equation} \label{eq:gradient_to_laplacian_identity}
    2 \partial_{x_\alpha} v(x) = \nabla_\mathbf{x}^2 [x_\alpha v(x)] - \frac{x_\alpha}{x} f''(x),
\end{equation}
where $\partial_{x_\alpha} \equiv \partial/\partial x_\alpha$ and $\nabla_\mathbf{x}^2 = \nabla_\mathbf{x} \cdot \nabla_\mathbf{x}$ ($\mathbf{x} \in \mathbb{R}^3$, $x = |\mathbf{x}|$).
Note here that $\mathbf{x}$ still corresponds to the electron-nucleus relative vector $\mathbf{x}_{iI} = \mathbf{r}_i - \mathbf{R}_I$ (omitting the particle indices for ease of notation), and $\nabla_\mathbf{x} = \nabla_{\mathbf{r}} = (\nabla_{\mathbf{r}_1}, \hdots, \nabla_{\mathbf{r}_N})$ is the configuration-space gradient.
We will use $\nabla_i \equiv \nabla_{\mathbf{r}_i}$.

\subsubsection{First IBP step: moving the gradient to the density} \label{sec:first_ibp_step}

The variance problem originates from the short-range divergence of the electron-nucleus interaction $v_{en}$.
To isolate the singular contribution, we split the interaction in a short and a long-range part (as in \cref{eq:v_en_sr_lr_split}), that is, $v_{en} = v_{en}^{sr} + v_{en}^{lr}$, where $v_{en}^{sr}(x) \sim 1/x$ as $x \to 0$.
We note that this decomposition is not restricted to periodic systems, as one may set $v_{en}^{sr}=v_{en}$ and $v_{en}^{lr}=0$ for open systems.
In either case, the short-range Hellmann-Feynman force on nucleus $I$ can be written as
\begin{align} 
    \mathbf{F}_{I,sr}^\mathrm{HF} &=-\int d^3\mathbf{u} \ \nabla_{\mathbf{R}_I} v_{en}^{sr}(\mathbf{R}_I-\mathbf{u})\rho(\mathbf{u})
    \\
    &=-\int d^3\mathbf{u} \ v_{en}^{sr}(\mathbf{u}-\mathbf{R}_I) \nabla_\mathbf{u} \rho(\mathbf{u}), \label{eq:sr_hf_force}
\end{align}
where we used $-\nabla_{\mathbf{R}_I} v_{en}^{sr}(\mathbf{R}_I - \mathbf{u}) = \nabla_{\mathbf{u}} v_{en}^{sr}(\mathbf{R}_I - \mathbf{u})$ and then integrated by parts; the boundary term vanishes both for open and periodic systems.
Using $\rho(\mathbf{u}) = \langle \sum_{i=1}^N \delta(\mathbf{u} - \mathbf{r}_i) \rangle$, its gradient can be written as 
\begin{equation}
    \nabla_\mathbf{u}\rho(\mathbf{u}) = 2 \sum_i \Big\langle  \delta(\mathbf{u}-\mathbf{r}_i) \nabla_{\mathbf{r}_i} \log |\Psi| \Big\rangle,
\end{equation}
where we again integrated by parts.
Inserting this identity into \cref{eq:sr_hf_force} gives the first IBP-based estimator
\begin{equation} \label{eq:ibp1_estimator}
    \mathbf{F}_{I,sr}^{\mathrm{IBP1}}(\mathbf{r}) = - 2 \sum_i v_{en}^{sr}(\mathbf{r}_i-\mathbf{R}_I) \nabla_{\mathbf{r}_i} \log |\Psi|,
\end{equation}
with $\mathbf{F}_{I,sr}^\mathrm{HF} = \langle \mathbf{F}_{I,sr}^{\mathrm{IBP1}} (\mathbf{r}) \rangle$.
Thanks to the electron-nucleus cusp condition $\Psi(\mathbf{x}) = \Psi(0)[1 - Z_I x + \hdots]$, for $x \to 0$, the gradient behaves as $\nabla_i \log |\Psi| = \nabla_i \Psi/\Psi = (\partial_x \Psi/\Psi) \nabla_i x \sim -Z_I \mathbf{x}/x$.
The IBP1 estimator therefore scales only as
\begin{equation} \label{eq:scaling_ibp1}
    \mathbf{F}_{I,sr}^{\mathrm{IBP1}}(\mathbf{r}) \sim v_{en}^{sr}(x) \nabla_i \log |\Psi| \sim \frac{\mathbf{x}}{x^2} \sim \mathcal{O}\left(\frac{1}{x}\right),
\end{equation}
compared to the naive estimator scaling of $\mathbf{x}/x^3 = \hat{\mathbf{x}}/x^2 = \mathcal{O}(1/x^2)$, near coalescence points $x = |\mathbf{r}_i - \mathbf{R}_I| \to 0$.
This estimator thus provides a straightforward way to partially cure the variance problem.

\subsubsection{Second IBP step: gradient-to-Laplacian identity} \label{sec:second_ibp_step}

To further reduce the fluctuations, one can trade the singular derivative of $v_{en}^{sr}$ for a Laplacian acting on a less singular quantity, using the identity given in \cref{eq:gradient_to_laplacian_identity}.
The derivation is identical for open and periodic systems; only the choice of $v_{en}^{sr}$ differs.
For example, under OBC, $v_{en}^{sr}(x) \sim 1/x$, while under PBC, a typical choice is $v_{en}^{sr}(\mathbf{x}) = \sum_\mathbf{T} \mathrm{erfc}(\kappa |\mathbf{x} + \mathbf{T}|) / |\mathbf{x} + \mathbf{T}|$, corresponding to the Ewald short-range kernel, where $\{\mathbf{T}\}$ is the set of lattice translation vectors and $\kappa > 0$ is the (fixed) Ewald screening parameter.
In the following, we keep the notation shortcut $v_{en}^{sr}(x_{iI})$, with the understanding that, in the periodic case, this denotes the periodized kernel evaluated at the relative vector $\mathbf{x}_{iI} = \mathbf{r}_i - \mathbf{R}_I$, including the implicit sum over translation vectors.

We start from the expression of the $\alpha$-component of the short-range HF force on nucleus $I$, 
\begin{equation}
    F_{I,sr,\alpha}^\mathrm{HF} = \sum_i \int d\mathbf{r} |\Psi|^2 \partial_{x_{iI,\alpha}} v_{en}^{sr}(x_{iI}),
\end{equation}
where we used that $- \partial_{R_{I\alpha}} = \partial_{r_{i\alpha}} = \partial_{x_{iI,\alpha}}$.
We can directly recover the IBP1 estimator given in \cref{eq:ibp1_estimator} by moving the partial derivative $\partial_{x_{iI,\alpha}}$ to the weight $|\Psi|^2$ using a single integration by parts.
Instead, applying the gradient-to-Laplacian identity given in \cref{eq:gradient_to_laplacian_identity}, with $v \equiv v_{en}^{sr}$ and $f(x) = x v_{en}^{sr}(x)$, we obtain
\begin{equation}
    F^\mathrm{HF}_{I,\alpha,sr} = \sum_i \! \left\langle \frac{\nabla_i^2 \! \left[x_{iI,\alpha} v^{sr}_{en}(x_{iI}) \right]}{2} - \frac{x_{iI,\alpha}}{2 x_{iI}} f''(x_{iI}) \! \right\rangle.
\end{equation}

Integrating by parts the Laplacian term (still with vanishing boundary term), 
we get the following improved estimator, that we label ``IBP2'',
\begin{align}
    F_{I\alpha}^{\mathrm{IBP2}}(\mathbf{r})
    &= -\sum_j \nabla_j Q_{I\alpha} \cdot \nabla_j \log|\Psi| - R_{I\alpha}, \label{eq:ibp2_estimator} \\
    Q_{I\alpha}(\mathbf{r}) &= \sum_i \frac{x_{iI,\alpha}}{x_{iI}} f(x_{iI}), \label{eq:ibp2_estimator_Q_function} \\
    R_{I\alpha}(\mathbf{r}) &= \frac{1}{2}\sum_i \frac{x_{iI,\alpha}}{x_{iI}} f''(x_{iI}),
\end{align}
so that $\mathbf{F}_{I,sr}^\mathrm{HF} = \langle \mathbf{F}_{I,sr}^{\mathrm{IBP2}} (\mathbf{r}) \rangle$, and the $\mathbf{Q}$-function defined is analogous to the one introduced by Assaraf and Caffarel \cite{2003_assaraf_caffarel_zero_variance_force}.
Since $\nabla_j \log |\Psi| \sim \hat{\mathbf{x}}$ (as demonstrated between \cref{eq:scaling_ibp1,eq:ibp1_estimator}), the only potentially singular piece arising from the first term of $\nabla_j \left[(x_{iI,\alpha} / x_{iI})  f(x_{iI}) \right] = f(x_{iI}) \nabla_j \left(x_{iI,\alpha} / x_{iI} \right) + (x_{iI,\alpha} / x_{iI}) f'(x_{iI}) \nabla_j x_{iI}$ drops out because it is orthogonal to the radial direction, that is, $f(x_{iI}) \nabla_j (x_{iI,\alpha} / x_{iI}) \cdot \hat{\mathbf{x}} = 0$, and $f(0)$ is finite.
As a result, unlike the IBP1 estimator reported in \cref{eq:ibp1_estimator}, the IBP2 estimator remains finite at coalescence points.

Note that in the case of OBC, $v_{en}^{sr}(x) \sim 1/x$ and $f(x) \sim 1$ implies that $f''(x) = 0$, so $R_{I\alpha} = 0$, and \cref{eq:ibp2_estimator} reduces to the Assaraf-Caffarel improved form \cite{2003_assaraf_caffarel_zero_variance_force} (see the \textit{supplementary material}, Sec. IIIA).

\subsection{Further improvements} \label{sec:further_improvements}

\begin{table*}[!t]
    \centering
    \caption{Summary of the main estimators used in this work. ``HF'' stands for Hellmann-Feynman.}
    \label{tab:estimator_summary}
    \small
    \renewcommand{\arraystretch}{1.05}
    \begin{tabular*}{\textwidth}{@{\extracolsep{\fill}}llll@{}}
        \hline\hline
        Observable & Estimator & Equation(s) & Role \\
        \hline
        Energy derivative & HF + Pulay decomposition & \cref{eq:dEdp} & Defines the HF/Pulay split. \\
        Pulay terms & Naive Pulay estimator & \cref{eq:pulay} & Baseline with heavy tails due to nodes. \\
        Pulay terms & Covariance Pulay estimator & \cref{eq:pulay_covariance} & Same mean with lower variance. \\
        Generic observable & Acceptance-weighted estimator $\widetilde{O}$ & \cref{eq:acceptance_trick_estimator} & Conditional accept/reject average. \\
        Pulay terms & Proposal-averaged estimator $\overline{O}$ & \cref{eq:def_proposal_averaged_estimator} & Full proposal averaging with finite variance. \\
        Pulay terms & Finite-$K$ proposal average $\overline{O}_K$ & \cref{eq:def_K_proposal_averaged_estimator} & Practical interpolation between $\widetilde{O}$ and $\overline{O}$. \\
        Pulay terms & Regularized acceptance estimator $\widetilde{O}_\epsilon$ & \cref{eq:hard_cutoff_regularized_pulay_pressure} & Cuts near-node outliers with controlled bias. \\
        Pressure & Virial pressure $P^V$ & \cref{eq:Pvirial} & Hamiltonian pressure contribution. \\
        Pressure & Pulay pressure $P^{wf}$ & \cref{eq:pulay_pressure} & Wave function response contribution. \\
        HF force & Bare electron-nucleus HF estimator & \cref{eq:F_Ien_HF_v1} & Reference HF force with infinite variance. \\
        HF force & IBP1 estimator & \cref{eq:ibp1_estimator} & Softens $1/x^2$ singularity to $1/x$. \\
        HF force & IBP2 estimator (with $Q_{I\alpha}$) & \cref{eq:ibp2_estimator} & Cancels the leading singularity entirely. \\
        Pair correlation & Improved $g(r)$ estimator & \cref{eq:improved_pair_correlation_function} & Lower-noise alternative to shell/binning. \\
        \hline\hline
    \end{tabular*}
\end{table*}

Assaraf and Caffarel \cite{1999_assaraf_caffarel_zero_var} introduced a general strategy to define renormalized VMC estimators by replacing the bare observable $O$ with a modified observable $O'$ that has the same expectation value over the Born distribution $\pi \propto |\Psi|^2$, but typically a substantially reduced variance.
Although they did not phrase their construction in terms of control variates, it can be viewed as a special case of a broader class of so-called Stein control variates, as we show in Sec. IIIA of the \textit{supplementary material}.
This broader Stein operator perspective originates from Stein's method for distributional approximation \cite{1972_stein_method}.

A key development in this field was the generator interpretation due to Barbour \cite{1988_barbour_poisson,1990_barbour_diffusion}, which observes that if $\mathcal{G}$ is the infinitesimal generator of an ergodic Markov process with stationary distribution $\pi$, then $\langle \mathcal{G} [f] \rangle_\pi = 0$ under mild regularity and boundary conditions, for some smooth test function $f$.
Choosing the overdamped Langevin diffusion as such a process leads to the standard Langevin Stein operator \cite{2015_gorham_mackey_derives_langevin_stein_operator,2017_gorham_mackey_development_langevin_stein_operator},
\begin{equation}
    \mathcal{L}_\pi[\mathbf{A}](\mathbf{r}) = \nabla \cdot \mathbf{A}(\mathbf{r}) + \mathbf{A}(\mathbf{r}) \cdot \nabla \log \pi(\mathbf{r}),
\end{equation}
which satisfies $\langle \mathcal{L}_\pi[\mathbf{A}] \rangle_\pi = 0$ under mild boundary conditions and for any sufficiently regular vector field $\mathbf{A}(\mathbf{r})$; here $\nabla \equiv \nabla_\mathbf{r}$ is the configuration-space gradient.
This identity (and related) underpins Stein-based control variates and control functional methods for variance reduction \cite{2017_oates_control_functionals,2022_si_control_variate_for_mc_methods}.
For our purposes, one can define $O' \equiv O + \mathcal{L}_\pi[\mathbf{A}]$ and choose $\mathbf{A}$ so that $\langle O' \rangle_\pi = \langle O \rangle_\pi$ and, importantly, $\mathrm{Var}(O') < \mathrm{Var}(O)$, as exemplified for atomic forces (under open boundary conditions) in Sec. IIIA of the \textit{supplementary material}.

This framework can be extended and generalized beyond forces \cite{2003_assaraf_caffarel_zero_variance_force} to structural observables such as the one-body density, $\rho(r) \propto \langle \sum_i \delta(r - r_i) \rangle$, and the pair correlation function, $g(r) \propto \langle \sum_{i<j} \delta(r - r_{ij}) \rangle$ \cite{2004_adib_jarzynski_unbiased_structural_estimators,2007_assaraf_improved_one_body_density,2007_toulouse_zero_var_bias_pair_density,2009_manolo_intracule_densities}, where $r_i = |\mathbf{r}_i|$ and $r_{ij} = |\mathbf{r}_i - \mathbf{r}_j|$.
The bare estimators for these quantities are defined by delta function operators, which are notoriously noisy because they probe rare events on thin shells, leading to a systematic bias when implemented in practice via binning.
The key step is to choose a vector field $\mathbf{A}$ that is truncated by a Heaviside step function $\theta(s - \xi(\mathbf{r}))$, for an appropriate constraint $\xi(\mathbf{r})$ (for instance $\xi \equiv r_i$ for $\rho(r)$, or $\xi \equiv r_{ij}$ for $g(r)$).
Applying the divergence theorem then yields an identity (as detailed in Sec. IIIB of the \textit{supplementary material}) that replaces the noisy integration over a shell with an integration over the ball $\{\xi \le s \}$, inevitably reducing the variance, though at a higher computational cost given the added drift term $\nabla \log \pi$.
This latter term is however easily calculated in VMC, where $\pi \propto |\Psi|^2$, and corresponds to the quantum force $\nabla \log \pi = 2 \nabla \log \Psi$.

It is worth noting that these ideas have also been developed in the context of classical NVT (classical Monte Carlo and molecular dynamics) simulations \cite{2004_adib_jarzynski_unbiased_structural_estimators}, where the Boltzmann distribution $\pi \propto e^{-\beta U}$ is being sampled, and $\nabla_i \log \pi = \beta \mathbf{F}_i^\mathrm{cl}$, where $\beta$ is the inverse temperature and $\mathbf{F}_i^\mathrm{cl}$ is the classical force on particle $i$.
The resulting force-based estimators for $\rho(r)$ and $g(r)$ have already been shown to lower the variance \cite{2013_borgis_improved_pair_distribution_and_densities,2020_rotenberg_improved_densities_rdf_with_forces}.
We will further exemplify in the Results section the variance reduction for $g(r)$ in VMC, using the following estimator (derived in Sec. IIIB of the \textit{supplementary material})
\begin{equation} \label{eq:improved_pair_correlation_function}
    g(r) \propto \left\langle \sum_{i<j} \theta(r - r_{ij}) \left(\frac{2}{r_{ij}} + \hat{\mathbf{r}}_{ij} \cdot (\mathbf{F}_i - \mathbf{F}_j) \right) \right\rangle,
\end{equation}
with $\mathbf{F}_i \equiv \nabla_i \log \Psi$ being the quantum force acting on particle $i$.

Finally, we provide in Sec. IIIC of the \textit{supplementary material} a systematic extension to higher-order correlation functions involving a product of multiple delta functions, for example, angular distribution functions and other multi-constraint structural observables.
Such quantities are often calculated in atomistic simulations \cite{2014_distasio_car_structure_of_liquid_water,2023_niu_qmc_database_hydrogen,2025_istas_liquid_liquid_hydrogen_transition,2025_chai_solid_hydrogen_beyond_bo,2026_smith_moire_wigner_crystal}, but the associated bare estimators become increasingly noisy the more delta functions (i.e. constraints) are added.
The multi-constraint identity we derive offers an unbiased, variance-reduced alternative that applies not only to VMC/DMC, but also to classical NVT simulations (though, obviously, still at a higher computational cost).

While several of the references mentioned in this section introduce improved estimators for specific observables (mainly for forces, densities and pair correlations), they do not provide a single, systematic formulation that unifies them.
In Sec. III of the \textit{supplementary material}, we present such a unified framework explicitly: codimension-0 recovers the familiar zero-mean control variate correction, codimension-1 yields the exact $\delta \to \theta$ conversion used for $\rho(r)$ and $g(r)$, and codimension-2 extends the construction to products of delta functions.
We suspect our formulation could be further generalized and applicable for other observables and cases.
For ease of reference, \cref{tab:estimator_summary} collects the estimators used below and links each one to its defining equation.

\section{Results} \label{sec:results}

\begin{figure}[b]
    \centering
    \includegraphics[width=\columnwidth]{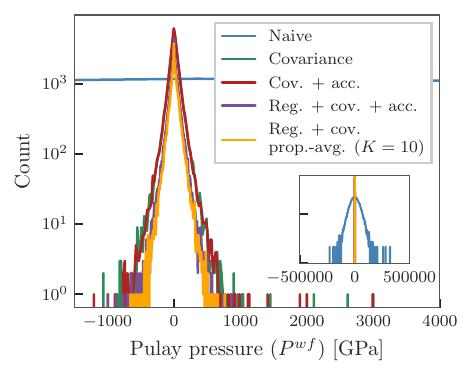}
    \caption{
    Comparison of Pulay pressure ($P^{wf}$) estimators for a periodic system of $N=128$ hydrogen atoms.
    The covariance estimator (``Cov.'') is given in \cref{eq:pulay_covariance}, the acceptance trick (``acc.'') in \cref{eq:acceptance_trick_estimator}, the finite proposal averaging (``prop.-avg.'') in \cref{eq:def_K_proposal_averaged_estimator}, and the hard-cutoff regularization (``reg.'') in
    \cref{eq:hard_cutoff_regularized_pulay_pressure}.
    A nodal cutoff of $\epsilon = 0.05 \ a_0$ was used for the regularization, and $K=10$ proposals were used for the proposal-averaged estimator.
    The empirical variances are reduced relative to the naive estimator by factors of $1.24\times 10^5$ (Cov.), $1.30\times 10^5$ (Cov. + acc.), $1.76\times 10^5$ (Reg + Cov. + acc.), and $1.95\times 10^5$ (Reg + Cov. + prop.-avg., $K=10$).
    The $99^\mathrm{th}$ and $99.9^\mathrm{th}$ percentiles of $|P^{wf}|$ decrease by $\sim 6\%$ and $\sim 14\%$, respectively, from $K=1$ to $K=10$
    }
    \label{fig:H_N128_rs119_comparing_pulay_estimators}
\end{figure}

We present VMC results for a periodic system of $N=128$ hydrogen atoms placed in a cubic cell, with the Hamiltonian taking the usual form given in \cref{eq:electronic_hamiltonian}.
We restrict our attention to a single high-pressure configuration at an electronic density of $r_s = 1.19$ (or roughly 650 GPa), where $r_s = a / a_0$, with $a$ and $a_0$ being the mean electronic distance and the Bohr radius, respectively.
The variational wave function used follows our earlier work \cite{2025_linteau_atomic_hydrogen}, with only minor modifications.

\begin{figure}[b]
    \centering
    \includegraphics[width=\columnwidth]{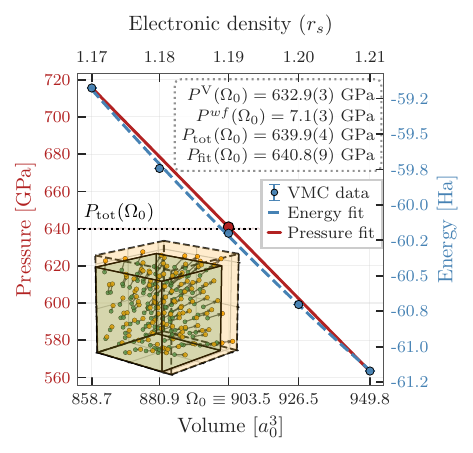}
    \caption{
    Comparison between the pressure obtained from the direct estimator and from a finite-difference fit.
    The direct estimate is computed as the sum of the Virial and Pulay contributions,
    $P_{\mathrm{tot}}(\Omega_0) = P^{\mathrm{V}}(\Omega_0) + P^{wf}(\Omega_0)$ (indicated with a dotted black line),
    where $\Omega_0$ is the simulation-cell volume of the original (unscaled) nuclear configuration.
    The finite-difference estimate, denoted $P_{\mathrm{fit}}(\Omega_0)$ (indicated by a large red marker), is obtained by differentiating a polynomial fit to the VMC
    energy data and evaluating it at $\Omega_0$.
    In the inset, the original and scaled cells are shown as light green and light orange surfaces, respectively, with solid and dashed lines marking their boundaries.
    The scaled configuration is illustrated by displaced nuclear positions; the arrows indicate geometric displacements (not forces).
    The inset is not to scale and is intended only as a schematic, since the actual volume changes considered are too small to be visible.
    }
    \label{fig:H_N128_rs119_pressure_vs_energy_vs_volume_vs_rs}
\end{figure}

We start by presenting results on the pressure.
As previously alluded to in \cref{sec:pressure}, the Virial pressure has finite variance and is not challenging to calculate precisely.
We therefore focus only on the Pulay-pressure contribution, given in \cref{eq:pulay_pressure}. 
In \cref{fig:H_N128_rs119_comparing_pulay_estimators}, we compare Pulay-pressure estimators built from covariance centering, acceptance weighting, regularization, and proposal averaging.
Because the associated sampled distributions are heavy-tailed and their second moment does not formally exist, as detailed in \cref{app:tail_exponents_moment_criterion_and_toy_model_example},
the variance-reduction factors reported in the caption of \cref{fig:H_N128_rs119_comparing_pulay_estimators} should be interpreted as finite-sample diagnostics rather than intrinsic variances of the limiting estimator distributions.
They nevertheless provide a direct comparison of the estimator fluctuations on the same Monte Carlo data.
The covariance form provides the dominant improvement over the naive estimator, while acceptance weighting gives an additional gain at essentially no extra cost.
Regularization further stabilizes the estimator, and combining it with proposal averaging ($K=10$) gives the most precise estimator among those tested.

We evaluate each estimator using the same $M = 2^{15}= 32 \, 768$ (initial) Monte Carlo configurations $\{\mathbf{r}_i^{(\ell)}\}_{\ell=1}^M$.
These configurations were generated from many parallel Markov chains with long decorrelation sweeps between them to limit autocorrelation.
We typically used at least 1024 Metropolis steps between retained configurations, and at most a few configurations were retained from each chain.
For each $r_i^{(\ell)}$, $K$ additional configurations are then sampled to evaluate the proposal-averaged estimators (here for $K=1$ and 10).
The cost of proposal averaging therefore grows approximately linearly with $K$, not only because of the increased sampling cost, but, more importantly, because of the additional local energy evaluations.
Here, and in the following, proposal averaging was performed based on the same all-electron Gaussian moves used in the Markov chain.

Note that we choose $\epsilon$ for the regularization as the largest value for which the regularized Pulay-pressure mean is statistically indistinguishable from the $\epsilon \to 0$ reference value (i.e. the residual bias is below the error bar), while still yielding a clear variance reduction (see Sec. ID of the \textit{supplementary material} for further details).
Here $\epsilon = 0.05 \ a_0$ was systematically chosen.

\begin{figure}[t]
    \centering
    \includegraphics[width=\columnwidth]{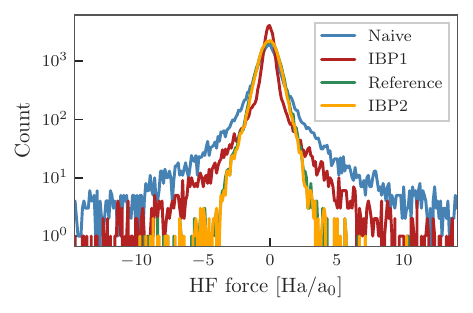}
    \caption{
    Comparison of the Hellmann-Feynman force estimators.
    The force component for the proton index 0 in the $x$-direction is shown without loss of generality.
    The integration-by-parts estimators ``IBP1'' and ``IBP2'' are presented in \cref{eq:ibp1_estimator,eq:ibp2_estimator}, respectively, while ``Reference'' denotes the estimator proposed in Ref.~\cite{2024_qian_space_warp_forces}, also described in Sec. IV of the \textit{supplementary material}.
    The empirical variances are reduced relative to the naive estimator by factors of $249\times$ (IBP1), $280\times$ (IBP1 + acc.), $1207\times$ (Reference), $1215\times$ (Reference + acc.), $1286\times$ (IBP2), and $1295\times$ (IBP2 + acc.), where ``+ acc.'' denotes the acceptance-weighted variant of the corresponding estimator (not shown for clarity).
    }
    \label{fig:H_N128_rs1.19_comparing_HF_estimators}
\end{figure}

To determine the proportion of the Pulay contribution to the total pressure, and simultaneously assess the accuracy of our pressure calculation, we compare with finite difference in \cref{fig:H_N128_rs119_pressure_vs_energy_vs_volume_vs_rs}.
We see that the pressure obtained by direct calculation $P_\mathrm{tot}(\Omega_0) = P^\mathrm{V}(\Omega_0) + P^{wf}(\Omega_0) = 639.9(4)$ GPa is in agreement with the pressure obtained with finite difference $P_\mathrm{fit}(\Omega_0) = 640.8(9)$ GPa, where $\Omega_0$ is the volume of the cell where the original (non-dilated) nuclear configuration is placed.
More importantly, although the NQS-based wave function is very good at modeling the ground state (as demonstrated in our earlier work \cite{2025_linteau_atomic_hydrogen}), the Pulay contribution $P^{wf}(\Omega_0) = 7.1(3)$ GPa is non-negligible.

We further note that the NQS-based orbitals have an explicit dependence on the nuclear coordinates $\mathbf{R}$ and on the cell geometry, so the dilation derivative (see \cref{eq:dilation_derivative}) naturally propagates through the orbitals.
This contrasts with the common practice of using density functional theory (DFT) orbitals where the $\mathbf{R}$-dependence is not explicit. There, the orbital response to changes in $\mathbf{R}$ enters implicitly through the
Kohn-Sham equations.
Neglecting their contributions in the Pulay term will introduce a systematic bias.

Similarly to pressure, atomic forces greatly benefit from variance-reduced estimators. 
We illustrate this on the same hydrogen system in \cref{fig:H_N128_rs1.19_comparing_HF_estimators}, where we overlay line distributions of several Hellmann-Feynman estimators ordered by decreasing variance (which in this case correlates to shrinking tails). 
The corresponding variance-reduction factors relative to the naive estimator are reported in the caption of \cref{fig:H_N128_rs1.19_comparing_HF_estimators}, consisting again of a finite sample diagnostics given the heavy-tailed nature of the distributions. 
The IBP1 estimator introduced in \cref{sec:first_ibp_step} already suppresses the number of outliers compared with the naive estimator.
This clearly shows how softening the near-coalescence divergence from $\mathcal{O}(1/x^2)$ to $\mathcal{O}(1/x)$ substantially improves the estimator's reliability.
The resolution of the Hellmann-Feynman estimator can be improved further by removing the leading singularity, as outlined in \cref{sec:second_ibp_step} with the IBP2 estimator, which yields the strongest tail suppression among the tested estimators.
For comparison, we also include an alternative estimator with a similar variance profile proposed by Qian \textit{et al.} \cite{2024_qian_space_warp_forces}.
Their construction extends the Assaraf-Caffarel approach \cite{2003_assaraf_caffarel_zero_variance_force} to periodic systems, following a more heuristic approach.
We describe the relation of this estimator to our framework presented in \cref{sec:ibp_for_hf_forces} and provide additional details in Sec. IV of the \textit{supplementary material}.
Finally, for all three estimators, applying the acceptance trick on top further reduces the number of outliers and therefore systematically lowers the variance (though to a lesser extent).

\begin{figure}[t]
    \centering
    \includegraphics[width=\columnwidth]{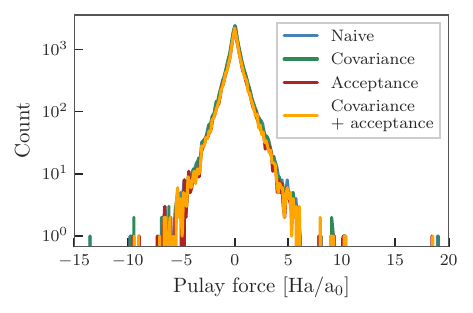}
    \caption{
    Comparison of the Pulay force estimators.
    The covariance estimator is given in \cref{eq:pulay_covariance}, and the acceptance trick is given in \cref{eq:acceptance_trick_estimator}.
    The force component for the proton index 0 in the $x$-direction is shown without loss of generality.
    }
    \label{fig:H_N128_rs1.19_comparing_pulay_force_estimators}
\end{figure}

\begin{figure*}[t]
    \centering
    \includegraphics[width=\textwidth]{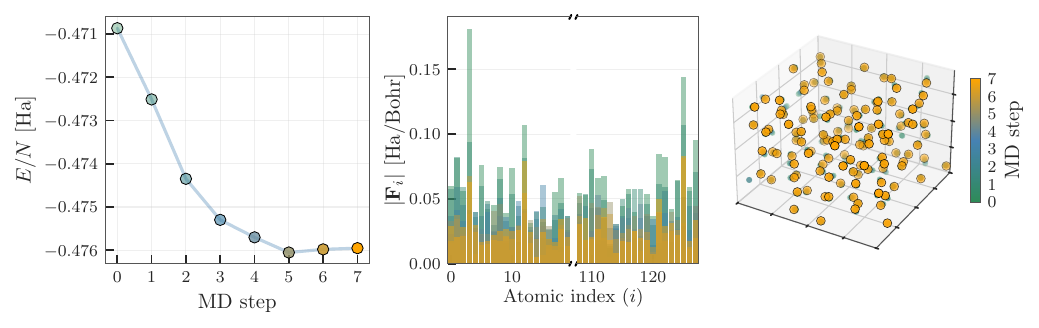}
    \caption{
    Molecular dynamics evolution of the hydrogen system, showing from left to right the electronic energy, the atomic forces, and the nuclear configurations.
    The molecular dynamics protocol consists of an iterative three-step procedure: 
    (1) a short VMC optimization;
    (2) the evaluation of the forces associated with the optimized wave function and nuclear configuration;
    and (3) the propagation of the nuclear coordinates using the damped velocity Verlet integration scheme based on these forces, after which the cycle is restarted with the updated coordinates.
    }
    \label{fig:md_evolution_n128_rs119}
\end{figure*}

In contrast to the Hellmann-Feynman force estimator, the naive Pulay force estimator is not noisy, as seen in \cref{fig:H_N128_rs1.19_comparing_pulay_force_estimators}.
While the Pulay force and Pulay pressure share the same structure, namely $F_{I\alpha}^{wf} \propto \langle \Delta E \partial_{R_{I\alpha}} \log \Psi^* \rangle$ and $P^{wf} \propto \langle \Delta E \mathcal{D} \log \Psi^* \rangle$, the usefulness of the covariance form depends on the scale of the log-derivative term multiplying $\Delta E$.
For pressure, the dilation operator $\mathcal{D}$ acts globally, resulting in an extensive $\mathcal{O}(N)$ contribution (i.e. the term $3N/2$ in \cref{eq:dlog_tilde_psi_to_dlogpsi}), so centering $\mathcal{D} \log \Psi^*$ removes a large and noisy zero-mean contribution and yields a strong variance reduction (as shown in \cref{fig:H_N128_rs119_comparing_pulay_estimators}).
Similarly, covariance centering can be very effective for calculating
parameter derivatives in optimization problems 
when $\partial_\theta \log \Psi^*$ contains normalization contributions that are extensive, given some variational parameter $\theta$.
By contrast, $\partial_{R_{I\alpha}} \log \Psi^*$ is a local response to displacing one nucleus, typically controlled by the environment around nucleus $I$.
As a result, it does not scale with system size so its removable mean part is much smaller.
This explains why the line distributions shown in \cref{fig:H_N128_rs1.19_comparing_pulay_force_estimators} are nearly indistinguishable, with no clear variance reduction gain and no further outlier suppression.

As an application of our force estimators, we perform a few molecular dynamics (MD) steps starting from the original hydrogen configuration, using a standard damped velocity Verlet integrator scheme.
We report the electronic energy, atomic forces, and nuclear positions along the trajectory in \cref{fig:md_evolution_n128_rs119}.
The MD loop follows a simple three-step protocol, as described in the caption.
The total force at each step includes both the Hellmann-Feynman IBP2 and Pulay contributions. 
Over only five MD steps, the energy per particle decreases by $\sim 5$ mHa, and the force magnitudes systematically decrease as the nuclei relax.

\begin{figure}[b]
    \centering
    \includegraphics[width=\columnwidth]{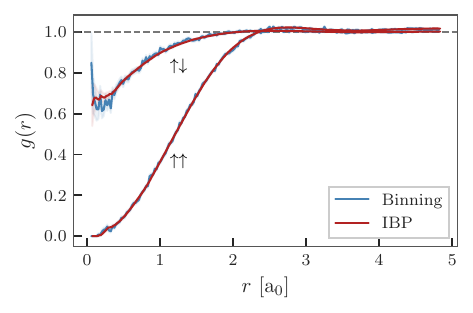}
    \caption{
    Spin-resolved pair correlation function with the naive binning and IBP estimators for a (relatively) small number of Monte Carlo configurations (i.e. 8192).
    The IBP estimator is defined in \cref{eq:improved_pair_correlation_function}.
    }
    \label{fig:H_N128_rs119_el-el-pair-correlation_nsamples8192_nchains1024}
\end{figure}

Because our wave function uses NQS-based orbitals with an explicit $\mathbf{R}$-dependence, we can employ transfer learning by initializing each new VMC optimization from the parameters obtained at the previous MD step.
As a result, only the first VMC optimization needs to be run in full.
Subsequent VMC optimizations typically converge within a few iterations and can be orders of magnitude faster, provided the MD step size is chosen sufficiently small.
This contrasts with the common use of fixed DFT orbitals, whose coefficients are tied to a specific nuclear geometry and are generally not transferable across configurations. 
As noted in the Introduction, this also motivates foundation model wave functions \cite{2024_scherbela_transferable_fermionic_neural_wf,2024_gao_neural_pfaffian,2025_rende_foundation_neural_network_quantum_states}, which could provide accurate forces across many geometries while avoiding re-optimization at each MD step.

As described in \cref{sec:further_improvements}, the integration by parts (IBP) framework extends naturally beyond forces and pressure to other observables.
In \cref{fig:H_N128_rs119_el-el-pair-correlation_nsamples8192_nchains1024}, we show the spin-resolved electron-electron pair correlation function of the hydrogen configuration, calculated using \cref{eq:improved_pair_correlation_function}.
For both parallel and anti-parallel spins, the IBP estimator is markedly less noisy than the standard binning estimator and yields a smooth and stable signal, which can generally be achieved with significantly fewer Monte Carlo samples.
Its statistical quality is also expected to improve as the radius $r$ increases, since the estimator averages over a ball of radius $r$, thus enhancing the signal-to-noise ratio.
This would be particularly advantageous for systems whose pair-correlation functions exhibit sharp features.

\section{Conclusion} \label{sec:conclusion}

We have described a set of simple and practical strategies to reduce the variance of force- and pressure-related estimators in variational Monte Carlo.
For Pulay-type contributions, we showed that the Metropolis acceptance trick weakens the nodal divergence from power-law to logarithmic, and that full proposal
averaging removes it entirely. 
For finite proposal averaging, additional regularization remains useful, with the resulting bias kept under control.
For Hellmann-Feynman forces, we derived compact integration-by-parts estimators that remove the Coulomb singularity while remaining straightforward to implement in both open and periodic systems.
Beyond their immediate relevance for quantum Monte Carlo calculations, such improved force estimators are also important in a broader context.
Indeed, accurate force calculations are increasingly being used as training data for machine-learned interatomic force fields, which in turn enable large-scale classical and path-integral molecular dynamics simulations \cite{2023_niu_qmc_database_hydrogen,2024_Ceperley,2023_huang_machine_learning_dmc_forces,2025_Tenti}.

We also showed that the same perspective extends beyond forces and pressure.
In particular, the divergence-theorem identities introduced here provide a route to variance-reduced estimators for other observables, including pair correlations and angular distribution functions, by using zero-mean control variates or replacing shell integrals (delta functions) with volume integrals (Heaviside step functions).

A key point is that all of these improvements are compatible with standard VMC codes and incur little additional computational cost.
These can be combined with flexible variational wave functions, such as NQS, to accurately compute forces and pressure for equations-of-state calculations and structural optimization.

Looking ahead, it would be natural to extend the same ideas to other observables involving wave function derivatives, such as higher-order response properties or the stress tensor in periodic systems
\cite{2021_nakano_forces_for_phonon_dispersion,2023_Chen,2023_zhang_eos_high_pressure_magnesium_oxide}. 
Combined with transferable or foundation model wave functions, these methods can be naturally extended to fixed-pressure calculations \cite{2025_linteau_helium4_2d,2025_chai_solid_hydrogen_beyond_bo},
enabling the minimization of the Gibbs free energy for determining atomic structures and cell geometries in materials.

While finalizing this manuscript, we became aware of a recent preprint \cite{2026_wan_blurred_sampling} that addresses related questions, in particular the heavy-tailed nodal outliers giving rise to infinite variance, and introduces the so-called \textit{blurred sampling} method.

\section*{Supplementary Material}

See the \textit{supplementary material} for detailed derivations of the near-node scaling of regularized Pulay estimators, of the gradient-to-Laplacian and divergence-theorem identities, and a description of a periodic alternative for the \(\mathbf{Q}\)-function.

\section*{Acknowledgments}
We thank D. Ceperley, C. Pierleoni, Y. Yang, K. Nakano, Z.-Q. Wan, S. Zhang, and G. Mazzola for useful discussions
and for providing reference data for comparisons.
The authors acknowledge support from SEFRI under Grant No.\ MB22.00051 (NEQS - Neural Quantum Simulation) and from the French Agency for Research, project SIX (ANR-23-CE30-0022).
This work was also supported by a grant from the Swiss National Supercomputing Centre (CSCS) under project ID a117 on Alps, as part of the Swiss AI Initiative.

\section*{Data Availability}

The data that supports the findings of this study are available from the corresponding author upon reasonable request.

\appendix

\section{Near-node scaling of acceptance-weighted, proposal-averaged, and regularized Pulay estimators}
\label[appendix]{app:near-node-scaling}

In this appendix, we analyze the near-node scaling of several estimators for Pulay-type observables.
We first introduce notation and assumptions used throughout this appendix, then derive the variance of the naive estimator, the acceptance-weighted estimator, and the proposal-averaged estimator.
We then summarize the asymptotic variance and bias scalings of the various improved estimators, and finally present two illustrative examples, namely a toy model and a time-dependent variational Monte Carlo (tVMC) application, that complement the dense-hydrogen example considered in the main text.

\subsection{Nodal surface preliminaries}

We consider an observable $O(\mathbf{r})$ sampled in VMC from the stationary distribution $\pi(\mathbf{r})=|\Psi(\mathbf{r})|^2$.
Throughout this appendix, $\mathbf{r}_i$ denotes the current configuration of the Markov chain and $\mathbf{r}_f$ the proposed configuration.
Expectations over $(\mathbf{r}_i,\mathbf{r}_f)$ are understood as
\begin{equation}
    \langle \cdots \rangle = \int d^3\mathbf{r}_i \pi(\mathbf{r}_i) \int d^3\mathbf{r}_f p(\mathbf{r}_f | \mathbf{r}_i) (\cdots),
\end{equation}
where $p(\mathbf{r}_f | \mathbf{r}_i)$ is the proposal distribution.
For simplicity, and without loss of generality for the present scaling arguments, we assume a symmetric proposal, $p(\mathbf{x} | \mathbf{y}) = p(\mathbf{y} | \mathbf{x})$, so that the Metropolis-Hastings acceptance probability reduces to
\begin{equation}
    a(\mathbf{r}_i\to \mathbf{r}_f) = \min \left(1, \frac{|\Psi(\mathbf{r}_f)|^2}{|\Psi(\mathbf{r}_i)|^2}\right).
\end{equation}
We also assume that $p(\mathbf{r}_f|\mathbf{r}_i)$ is a normalized, local, symmetric proposal kernel that remains finite and non-zero in the vicinity of the nodal surface.
Here normalized means that $p$ is a proper conditional probability density (i.e. $\int d^3\mathbf{r}_f p(\mathbf{r}_f|\mathbf{r}_i)=1$), local means that the proposal scale is fixed and does not shrink with the distance to the node or with the cutoff $\xi_c$ defined below, and symmetric means that $p(\mathbf{x}|\mathbf{y}) = p(\mathbf{y}|\mathbf{x})$, so that the
Metropolis-Hastings ratio reduces to the ratio of Born weights.
Under these assumptions, the proposal density only contributes finite pre-factors to the asymptotic scalings presented below.
We denote the nodal surface as 
\begin{equation}
    \mathcal{N} = \{\mathbf{r} \in \mathbb{R}^{3N} | \Psi(\mathbf{r}) = 0 \}.
\end{equation}
Close to a generic point on $\mathcal{N}$, we introduce local coordinates $\mathbf{r} = (\xi, \mathbf{r}_{\|})$, where $\xi \in \mathbb{R}$ is the signed distance to the node and $\mathbf{r}_{\|} \in \mathbb{R}^{3N-1}$ parameterizes the nodal surface.
Assuming the derivative along the normal direction does not vanish, i.e. $\Phi(\mathbf{r}_{\|}) \equiv \partial_\xi \Psi(\xi, \mathbf{r}_{\|})|_{\xi=0} \neq 0$, a first-order Taylor expansion along that direction gives
\begin{equation}
    \Psi(\xi,\mathbf{r}_{\|}) = \Phi(\mathbf{r}_{\|}) \xi + \mathcal{O}(\xi^2),
\end{equation}
so that
\begin{equation}
    |\Psi(\xi, \mathbf{r}_{\|})|^2 = |\Phi(\mathbf{r}_{\|})|^2 \xi^2 + \mathcal{O}(|\xi|^3) \sim \xi^2.
\end{equation}
For Pulay-type observables,
\begin{equation}
    O(\mathbf{r})\equiv (E_L(\mathbf{r})-E) \partial_p \log \Psi^*(\mathbf{r}),
\end{equation}
one generically has near the node
\begin{equation}
    E_L(\mathbf{r})\sim \xi^{-1},
    \ \
    \partial_p \log \Psi^*(\mathbf{r})\sim \xi^{-1},
    \ \
    O(\mathbf{r})\sim \xi^{-2}.
\end{equation}

Since we are only interested in the leading near-node scaling, and the two sides of the nodal surface differ only by finite prefactors, we restrict the analysis to $\xi > 0$.
The tangential integrals over $\mathbf{r}_{\|}$ remain finite and only contribute overall constants, so from now on we keep track only of the normal coordinate $\xi > 0$.
We denote by $\xi_0$ a fixed microscopic scale delimiting the region where the nodal expansion is valid, and by $\xi_c$ (with $\xi_c \ll \xi_0$) a formal short-distance cutoff used only to display the divergence of unregularized estimators.

\subsection{Naive estimator: power-law divergent variance}

For the naive estimator $O(\mathbf{r}) \sim \xi^{-2}$, the mean remains finite because
\begin{equation}
    \langle O\rangle \sim \int_0^{\xi_0} d\xi  |\Psi|^2 O \sim \int_0^{\xi_0} d\xi \xi^2 \xi^{-2},
\end{equation}
which is integrable at $\xi=0$.
By contrast, the second moment behaves as
\begin{equation}
    \langle O^2 \rangle \sim \int_{\xi_c}^{\xi_0} d\xi  \xi^2 \xi^{-4} = \int_{\xi_c}^{\xi_0} d\xi \xi^{-2} = \frac{1}{\xi_c} - \frac{1}{\xi_0} \sim \xi_c^{-1}.
\end{equation}
Therefore the naive Pulay estimator has a power-law divergent variance,
\begin{equation}
    \boxed{
    \mathrm{Var}(O)\sim \xi_c^{-1}
    }
    \ .
\end{equation}

\subsection{Acceptance-weighted estimator}

The acceptance trick, described in \cref{sec:acceptance_trick}, replaces the naive one-point estimator $O(\mathbf{r})$ by the conditional two-point estimator
\begin{equation} 
    \widetilde{O}(\mathbf{r}_i, \mathbf{r}_f) = a O(\mathbf{r}_f) + (1 - a) O(\mathbf{r}_i).
\end{equation}
This estimator is unbiased, as previously discussed in the main manuscript in the ``Acceptance trick'' section.

Near the node, the acceptance probability scales as $a(\mathbf{r}_i \to \mathbf{r}_f) \sim \min (1, \xi_f^2 / \xi_i^2)$.
It is therefore natural to split the $(\xi_i,\xi_f)$ plane into the two regions
\begin{equation} \label{eq:two_nodal_regions}
    \text{(A)} \quad \xi_f \ge \xi_i, \qquad \text{(B)} \quad \xi_f < \xi_i.
\end{equation}
In region (A), one has $a = 1$, so $\widetilde{O} \sim O(\xi_f) \sim \xi_f^{-2}$ and therefore
\begin{equation}
    \langle \widetilde{O}^2 \rangle_A \sim \int_{\xi_c}^{\xi_0} d\xi_i \xi_i^2 \int_{\xi_i}^{\xi_0} d\xi_f \xi_f^{-4} \sim \int_{\xi_c}^{\xi_0} d\xi_i \xi_i^{-1}.
\end{equation}
In region (B), one has $a \sim \xi_f^2/\xi_i^2$, hence $a O(\xi_f) \sim \xi_i^{-2}$ and $(1-a) O(\xi_i) \sim \xi_i^{-2}$, so $\widetilde{O} \sim \xi_i^{-2}$.
This gives
\begin{equation}
    \langle \widetilde{O}^2 \rangle_B \sim \int_{\xi_c}^{\xi_0} d\xi_i \xi_i^2 \int_{0}^{\xi_i} d\xi_f \xi_i^{-4} \sim \int_{\xi_c}^{\xi_0} d\xi_i \xi_i^{-1}.
\end{equation}
In both regions, we obtain the logarithmic divergence
\begin{equation} \label{eq:log_scaling_of_acceptance_weighted_estimator}
    \boxed{
    \mathrm{Var}(\widetilde{O}) \sim \log \left(\frac{\xi_0}{\xi_c}\right)
    }
    \ .
\end{equation}
Thus the acceptance trick weakens the variance divergence from power-law to logarithmic.

\subsubsection{Rao-Blackwell interpretation} \label[appendix]{app:rao-blackwell_interpretation}

In this section, we make explicit the variance-reduction mechanism behind the acceptance-weighted estimator.
We first derive the variance inequality stated in the main text in \cref{eq:rao_blackwell_O_tilde_vs_O}, that is,
\begin{equation} 
    \mathrm{Var}( \widetilde{O}(\mathbf{r}_i,\mathbf{r}_f) ) \le \mathrm{Var}(O(\mathbf{r}_{i+1})).
\end{equation}
For this, let
\begin{equation}
    X = A O(\mathbf{r}_f) + (1 - A) O(\mathbf{r}_i)
\end{equation}
denote the naive post-acceptance estimator obtained after the Metropolis accept/reject step, where $A \in \{0,1\}$ is the Metropolis acceptance indicator and $A \sim \mathrm{Bernoulli}(a)$.
At fixed $(\mathbf{r}_i,\mathbf{r}_f)$, the only remaining randomness is coming from the Bernoulli accept/reject variable $A$.
The acceptance-weighted estimator is obtained by averaging over this Bernoulli variable while keeping the proposed pair fixed, that is,
\begin{equation} \label{eq:rao_blackwell_conditioning}
    \widetilde{O}(\mathbf{r}_i, \mathbf{r}_f) = \mathbb{E}_\mathrm{Bern}[X|\mathbf{r}_i,\mathbf{r}_f].
\end{equation}
Thus the relevant comparison is between the conditional second moment of $X$ and the square of its conditional mean $\widetilde{O}$.
Since, at fixed $(\mathbf{r}_i,\mathbf{r}_f)$, only $A$ is random, we can use $A^2 = A$, $A(1-A)=0$ and $\mathbb{E}_\mathrm{Bern}[A|\mathbf{r}_i,\mathbf{r}_f] = a$ to obtain
\begin{equation}
    \mathbb{E}_\mathrm{Bern}[X^2|\mathbf{r}_i,\mathbf{r}_f] 
    = a O(\mathbf{r}_f)^2 + (1-a) O(\mathbf{r}_i)^2.
\end{equation}
By contrast,
\begin{equation} 
    \widetilde{O}(\mathbf{r}_i, \mathbf{r}_f)^2 = [a O(\mathbf{r}_f) + (1-a) O(\mathbf{r}_i)]^2.
\end{equation}
Their difference is
\begin{align} \label{eq:local_rao_blackwell_O_tilde_vs_O}
    \mathbb{E}_\mathrm{Bern}[X^2 &| \mathbf{r}_i,\mathbf{r}_f] - \widetilde{O}(\mathbf{r}_i, \mathbf{r}_f)^2 \nonumber\\
    &= a(1-a)[O(\mathbf{r}_f) - O(\mathbf{r}_i)]^2 \ge 0.
\end{align}
This is the local form of the Rao-Blackwell variance reduction, where the acceptance-weighted estimator removes the conditional variance associated with the Bernoulli accept/reject decision.
Averaging this (local) identity over proposals gives the global variance inequality stated in the main text in \cref{eq:rao_blackwell_O_tilde_vs_O}, i.e. $\mathrm{Var}( \widetilde{O}(\mathbf{r}_i,\mathbf{r}_f) ) \le \mathrm{Var}(O(\mathbf{r}_{i+1}))$.
Alternatively, with the definition of $\widetilde{O}$ in \cref{eq:rao_blackwell_conditioning}, the law of total variance specifies that
\begin{equation}
    \mathrm{Var}(X) = \mathbb{E}[\mathrm{Var}(X|\mathbf{r}_i,\mathbf{r}_f)] + \mathrm{Var}(\widetilde{O}(\mathbf{r}_i,\mathbf{r}_f)),
\end{equation}
and since the first term is non-negative, \cref{eq:rao_blackwell_O_tilde_vs_O} follows.

We now apply this local identity to the near-node scaling.
In the region where the proposed point is farther from the node, $\xi_f \ge \xi_i$, one has $a = 1$ (i.e. region (A), as defined in \cref{eq:two_nodal_regions}).
The move is therefore accepted deterministically, so $X = \widetilde{O} = O$ pointwise in this region, and there is no conditional ``Bernoulli variance'' to remove.

The non-trivial reduction occurs when the proposal is closer to the node, $\xi_f < \xi_i$, so $0 < a < 1$ (i.e. region (B)).
In this region, $a \sim \xi_f^2 / \xi_i^2$, and the naive post-acceptance estimator has conditional second moment
\begin{equation}
    \mathbb{E}_\mathrm{Bern}[X^2|\xi_i,\xi_f] \sim \xi_i^{-4} + \xi_i^{-2} \xi_f^{-2}.
\end{equation}
The second term remains singular as $\xi_f \to 0$.
On the other hand, the acceptance-weighted estimator satisfies
\begin{equation}
    \widetilde{O}(\xi_i, \xi_f)^2 = (\mathbb{E}_\mathrm{Bern}[X|\xi_i, \xi_f])^2 \sim \xi_i^{-4},
\end{equation}
with no remaining $\xi_f^{-2}$ singularity.
The acceptance-weighted estimator can therefore be viewed as a Rao-Blackwell conditional estimator removing variance coming from the Bernoulli proposal (by averaging over it), more specifically removing a ``risky'' accepted contribution scaling as $\xi_i^{-2} \xi_f^{-2}$.
The logarithmic scaling, derived in \cref{eq:log_scaling_of_acceptance_weighted_estimator}, was then obtained by integrating $\widetilde{O}^2$ over the two regions of the joint proposal space.

\subsection{Proposal-averaged estimator} \label[appendix]{app:proposal_averaged_estimator}

Starting from the acceptance-weighted estimator $\widetilde{O}$, we define the proposal-averaged estimator 
\begin{equation} \label{eq:app_def_proposal_averaged_estimator}
    \overline{O}(\mathbf{r}_i) = \mathbb{E}[\widetilde{O}(\mathbf{r}_i,\mathbf{r}_f)|\mathbf{r}_i] = \int d^3\mathbf{r}_f p(\mathbf{r}_f|\mathbf{r}_i) \widetilde{O}(\mathbf{r}_i,\mathbf{r}_f). 
\end{equation}
This estimator is unbiased,
\begin{equation}
    \mathbb{E}[\overline{O}] 
    = \mathbb{E} \left[\mathbb{E}[\widetilde{O} | \mathbf{r}_i] \right]
    = \mathbb{E}[\widetilde{O}]
    = \mathbb{E}[O(\mathbf{r}_{i+1})]
    = \langle O \rangle,
\end{equation}
using the law of total expectation for the second equality.
Applying the law of total variance to $\widetilde{O}$ conditioned on $\mathbf{r}_i$,
\begin{align} \label{eq:var_proposal_averaged_estimator}
    \mathrm{Var}(\widetilde{O}) 
    &= \mathbb{E}[\mathrm{Var}(\widetilde{O}|\mathbf{r}_i)] + \mathrm{Var}(\mathbb{E}[\widetilde{O}|\mathbf{r}_i]) \nonumber\\
    &= \mathbb{E}[\mathrm{Var}(\widetilde{O}|\mathbf{r}_i)] + \mathrm{Var}(\overline{O}),
\end{align}
we find that $\overline{O}$ is a stricter conditioned estimator than the acceptance one,
\begin{equation}
    \mathrm{Var}(\overline{O}) \le \mathrm{Var}(\widetilde{O}) \le \mathrm{Var}(O(\mathbf{r}_{i+1})).
\end{equation}
To study the scaling behavior of $\widetilde{O}$, we consider 
\begin{equation}
    \overline{O}(\xi_i) \sim \int_0^{\xi_0} d\xi_f p(\xi_f | \xi_i)
    \Big[ a O(\xi_f) + (1 - a) O(\xi_i) \Big].
\end{equation}
In regions (A) and (B), as defined in the previous section in \cref{eq:two_nodal_regions}, we find respectively
\begin{align} 
    \overline{O}_A(\xi_i) &\sim \int_{\xi_i}^{\xi_0} d\xi_f \xi_f^{-2} \sim \xi_i^{-1}, \quad (\xi_f \ge \xi_i), \label{eq:prop_avg_scaling_after_integrating_prop_A} \\
    \overline{O}_B(\xi_i) &\sim \int_0^{\xi_i} d\xi_f \xi_i^{-2} \sim \xi_i^{-1}, \quad (\xi_f < \xi_i), \label{eq:prop_avg_scaling_after_integrating_prop_B}
\end{align}
so $\overline{O}(\xi_i) \sim \xi_i^{-1}$.   
The second moment then scales as
\begin{equation}
    \langle \overline{O}^2\rangle 
    \sim \int_0^{\xi_0} d\xi_i |\Psi|^2 \overline{O}(\xi_i)^2
    \sim \int_0^{\xi_0} d\xi_i \xi_i^2 \xi_i^{-2} 
    \sim \int_0^{\xi_0} d\xi_i,
\end{equation}
which is finite. 
Hence,
\begin{equation}\label{eq:finite_var_proposal_averaged_estimator}
    \boxed{\mathrm{Var}(\overline{O})=\mathcal{O}(1)}
\end{equation}
with no nodal divergence.
We have therefore constructed an unbiased estimator with finite variance.

\subsubsection{Finite-K approximation} \label[appendix]{app:K_proposal_averaged_estimator}

In practice, one may approximate \cref{eq:app_def_proposal_averaged_estimator} by drawing $K$ proposals
$\mathbf{r}_f^{(k)} \sim p(\cdot | \mathbf{r}_i)$ and using
\begin{equation}\label{eq:app_def_K_proposal_averaged_estimator}
    \overline{O}_K(\mathbf{r}_i) = \frac{1}{K}\sum_{k=1}^K \widetilde{O}(\mathbf{r}_i,\mathbf{r}_f^{(k)}).
\end{equation}
This estimator is unbiased since
\begin{equation}
    \mathbb{E}[\overline{O}_K | \mathbf{r}_i] 
    = \frac{1}{K} \sum_{k=1}^K \mathbb{E}[\widetilde{O}_k | \mathbf{r}_i] 
    = \frac{1}{K} \sum_{k=1}^K \overline{O} (\mathbf{r}_i) 
    = \overline{O}(\mathbf{r}_i),
\end{equation}
where we defined $\widetilde{O}_k \equiv \widetilde{O}(\mathbf{r}_i, \mathbf{r}_f^{(k)})$.
The variance then reads
\begin{align}
    \mathrm{Var}(\overline{O}_K | \mathbf{r}_i)
    &= \frac{1}{K^2} \sum_{k=1}^K \mathrm{Var}(\widetilde{O}_k | \mathbf{r}_i)
    + \frac{2}{K^2} \sum_{k<\ell} \mathrm{Cov}(\widetilde{O}_k, \widetilde{O}_\ell |  \mathbf{r}_i) \nonumber\\
    &= \frac{1}{K} \mathrm{Var}(\widetilde{O} | \mathbf{r}_i),
\end{align}
using the conditional independence of the proposals (and the variance formula for a linear combination of random variables). 
Applying again the law of total variance, but this time on $\overline{O}_K$ conditioned on $\mathbf{r}_i$, we get
\begin{align}\label{eq:var_K_proposal_averaged_estimator}
    \mathrm{Var}(\overline{O}_K) 
    &= \mathbb{E} \left[\mathrm{Var}(\overline{O}_K | \mathbf{r}_i) \right] + \mathrm{Var} \left(\mathbb{E}[\overline{O}_K | \mathbf{r}_i] \right) \nonumber\\
    &= \frac{1}{K} \mathbb{E} \left[\mathrm{Var}(\widetilde{O} | \mathbf{r}_i) \right] + \mathrm{Var}(\overline{O}).
\end{align}
We can re-arrange \cref{eq:var_proposal_averaged_estimator} into $\mathbb{E}[\mathrm{Var}(\widetilde{O}|\mathbf{r}_i)] = \mathrm{Var}(\widetilde{O}) - \mathrm{Var}(\overline{O})$,
and substitute into \cref{eq:var_K_proposal_averaged_estimator} to find
\begin{equation}
    \boxed{
    \mathrm{Var}(\overline{O}_K) = \left(1 - \frac{1}{K} \right) \mathrm{Var}(\overline{O}) + \frac{1}{K} \mathrm{Var}(\widetilde{O})
    }
    \ .
\end{equation}
Increasing $K$ continuously interpolates between the acceptance-weighted estimator ($K=1$) and the fully proposal-averaged one ($K\to\infty$).
As consistency checks, one recovers $\mathrm{Var}(\overline{O}_1)=\mathrm{Var}(\widetilde{O})$ and $\mathrm{Var}(\overline{O}_K)\to \mathrm{Var}(\overline{O})$ as $K\to\infty$.
The second term is the residual noise from incomplete proposal averaging.
Without additional regularization, this residual term inherits the same near-node logarithmic divergence as the acceptance estimator.

\subsubsection{Infinite-K limit}
Let us allow $K$ in \cref{eq:app_def_K_proposal_averaged_estimator} to depend on the current configuration $\mathbf{r}_i$. In particular, one may choose $K(\mathbf{r}_i)\sim \xi_i^{-\alpha}$ with $\alpha>0$, so that the number of proposals diverges ($K \to \infty$) as the nodal surface is approached. 
This yields a finite-variance estimator while avoiding the uniformly high computational cost of using a large constant $K$, independent of $\mathbf{r}_i$, throughout configuration space.

\subsection{Summary of asymptotic scalings}

\cref{tab:estimators_variance_bias} summarizes the asymptotic scalings of the variance and bias for the estimators discussed above, together with the regularized estimators $\widetilde{O}_\epsilon^{(1)}$, $\widetilde{O}_\epsilon^{(2)}$, and $\widehat{O}_\epsilon$ introduced in the \textit{supplementary material}.

\begin{table}[h]
\centering
\begin{tabular}{c|c|c}
estimator & variance & bias \\ \hline
$O(\mathbf{r})$ & $\xi_c^{-1}$ & $0$ \\
$\widetilde{O}$ & $\log(\xi_0/\xi_c)$ & $0$ \\
$\overline{O}$ & $\mathcal{O}(1)$ & $0$ \\
$\overline{O}_K$ & $(1-1/K)\,\mathcal{O}(1) + (1/K)\log(\xi_0/\xi_c)$ & $0$ \\
$\widetilde{O}_\epsilon^{(1)}, \widetilde{O}_\epsilon^{(2)}$ & $\log(\xi_0/\epsilon)$ & $\mathcal{O}(\epsilon^2)$ \\
$\widehat{O}_\epsilon = \chi_\epsilon \widetilde{O}$ & $\log(\xi_0/\epsilon)$ & $\mathcal{O}(\epsilon^{m+2})$
\end{tabular}
\caption{Bias–variance tradeoff across estimators}
\label{tab:estimators_variance_bias}
\end{table}

Recall that we denote by $\xi_0$ a fixed microscopic scale delimiting the region where the nodal expansion is valid, and by $\xi_c$ (with $\xi_c \ll \xi_0$) a formal short-distance cutoff used only to display the divergence of unregularized estimators. For the regularized estimators, a simple regularization scheme is to exclude a small region of width $\epsilon$ around the nodes. In addition, $\chi_\epsilon$ is the smooth cutoff function, $m$ denotes the number of canceled bias moments of $\chi_\epsilon$, and $K$ is the number of proposals used in the finite-$K$ proposal average.

\subsection{Toy model example} \label[appendix]{app:toy_model_example}

To complement the asymptotic near-node analysis, it is useful to consider a minimal two-dimensional toy model in which the nodal geometry is controlled by a single parameter.
We therefore consider one particle in two dimensions described by the variational wave function
\begin{equation}
    \psi_\theta(x,y) = e^{-2x^2-y^2} (x \sin\theta + y \cos\theta).
\end{equation}
In this toy model, the particle is not subject to any external confinement (i.e. the potential vanishes) so the Hamiltonian contains only the kinetic energy operator, $H=-\nabla^2$.
The Gaussian factor localizes the particle and keeps the wave function normalizable, while the linear factor defines a straight nodal line, given by $y = -x \tan\theta$ when $\cos\theta \neq 0$, so that $\theta$ simply rotates the node in the plane.

The quantity of interest is the derivative of the variational energy,
\begin{equation}
    F_\theta = \frac{dE(\theta)}{d\theta} = 2 (E_L - \langle H \rangle ) \frac{\partial \log \psi_\theta}{\partial \theta},
\end{equation}
which is not a physical force since $\theta$ is a variational parameter rather than a coordinate of the Hamiltonian, but it provides a Pulay benchmark for the same near-node statistical issues that appear in physical force estimators, given that $E_L \sim \xi^{-1}$ and $F_\theta \sim \xi^{-2}$, where $\xi = x \sin \theta + y \cos \theta$.

\begin{figure}[t]
  \centering
  \includegraphics[width=0.5\textwidth]{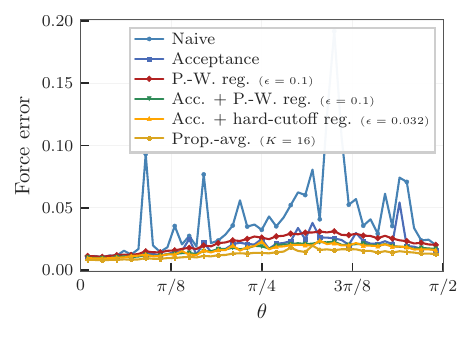}
  \caption{
  Statistical error of the toy model force $F_\theta$ as a function of $\theta$ for the Gaussian state $\psi_\theta(x,y)=e^{-2x^2-y^2}
  (x\sin\theta+y\cos\theta)$. The curves compare the naive, acceptance-weighted (or ``Acc.'' for short) [\cref{eq:acceptance_trick_estimator}], finite-$K$ proposal-averaged (or Prop.-avg.) [\cref{eq:def_K_proposal_averaged_estimator}], hard-cutoff regularized (or ``reg.'') [\cref{eq:hard_cutoff_regularized_pulay_pressure}] estimators, as well as the extrapolation scheme from Pathak and Wagner \cite{2020_pathak_wagner_gradient_regularization} (denoted ``P.-W.'').
  }
  \label{fig:toy_model_force_estimators}
\end{figure}

As seen in \cref{fig:toy_model_force_estimators}, this Gaussian toy model clearly exposes the variance reduction obtained from the various estimators proposed in this work.
It shows for instance how proposal averaging $(K > 1)$ improves upon the acceptance-weighted estimator $(K=1)$, and how regularizations can be used to further stabilize the results.

\subsection{Tail exponents, moment criterion, and toy model example} \label[appendix]{app:tail_exponents_moment_criterion_and_toy_model_example}

We formalize in this section how the asymptotic complementary cumulative distribution function (CCDF) exponent is obtained from the near-node scaling of a generic Pulay-type observable.
We then show how this exponent, denoted $\alpha$, determines whether the $n$-th moment of each estimator distribution is finite.
The toy model will then serve as an example to visualize the tail scaling of the variance-reduced estimators.

\begin{figure*}
    \centering
    \includegraphics[width=0.95\textwidth]{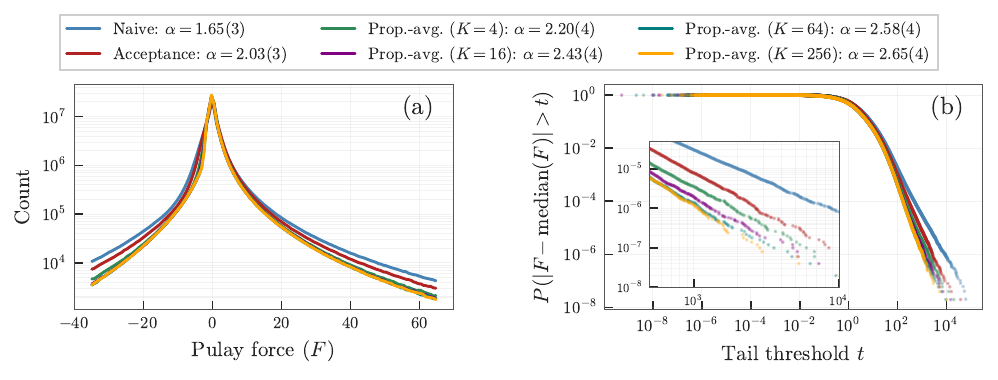}
    \caption{
    Tail analysis for the toy model Pulay force estimators (introduced in \cref{app:toy_model_example}) at $\theta = 0.25$ using $10^8$ Monte Carlo samples.
    (a) Force estimator distributions for naive, acceptance-weighted ($K=1$), and proposal-averaged estimators (for $K=4, 16, 64$ and 256).
    (b) For each estimator, the right-tail CCDF is shown together with the corresponding tail exponent estimate $\alpha$, reported in the legend. 
    The exponents were obtained using the Hill tail-index estimator \cite{1975_hill_tail_of_distribution,1996_beirlant_tail_index_estimation} with uncertainties provided by the bootstrap method \cite{1986_efron_bootstrap_methods} as implemented in the TREAT code \cite{2019_lopez_conduit_tail_regression_qmc,2019_lopez_treat_code_github}.
    }
    \label{fig:toy_model_force_distribution_and_ccdf}
\end{figure*}

For a random variable $X$, we define the CCDF as 
\begin{equation}
    \mathrm{Pr}(|X| > t) = \left\langle \mathbf{1}_{\{|X| > t \}} \right\rangle,
\end{equation}
where $\mathbf{1}$ is the indicator function and $t$ is some fixed tail threshold.
Let us focus on the case of interest, where $X$ is the naive force estimator, denoted $F$, which, in terms of the near-node normal coordinate $\xi$, can be written as
\begin{equation}
    \mathrm{Pr}(|F| > t) \sim \int_0^{\xi_0} d\xi |\Psi(\xi)|^2 \mathbf{1}_{\{|F(\xi)| > t \}}.
\end{equation}
For the naive estimator, $F \sim \xi^{-2}$ so the event $|F| > t$ is equivalent to  $\xi < t^{-1/2}$ and 
\begin{equation}
    \mathrm{Pr}(|F| > t) \sim \int_0^{t^{-1/2}} d\xi \, \xi^2 \sim t^{-3/2}.
\end{equation}
For the acceptance-weighted estimator $\widetilde{F}(\mathbf{r}_i, \mathbf{r}_f)$, in region (A) ($\xi_f \ge \xi_i$), one has $\widetilde{F} \sim \xi_f^{-2}$, hence
\begin{equation}
    \mathrm{Pr}_A(|\widetilde{F}| > t) \sim \int_0^{t^{-1/2}} d\xi_f \int_0^{\xi_f} d\xi_i \xi_i^2 \sim t^{-2}.
\end{equation}
Similarly, in region (B) ($\xi_f < \xi_i$), $\widetilde{F} \sim \xi_i^{-2}$ so
\begin{equation}
    \mathrm{Pr}_B(|\widetilde{F}| > t) \sim \int_0^{t^{-1/2}} d\xi_i \xi_i^2 \int_0^{\xi_i} d\xi_f \sim t^{-2},
\end{equation}
and therefore $\mathrm{Pr}(|\widetilde{F}| > t) \sim t^{-2}$.
For the proposal-averaged estimator $\overline{F}(\xi_i)$, the proposal coordinate has already been integrated out and $\overline{F} \sim \xi_i^{-1}$ (see \cref{eq:prop_avg_scaling_after_integrating_prop_A,eq:prop_avg_scaling_after_integrating_prop_B}), so $|\overline{F}| > t$ implies $\xi_i < t^{-1}$ and 
\begin{equation}
    \mathrm{Pr}(|\overline{F}| > t) \sim \int_0^{t^{-1}} d\xi_i \xi_i^{2} \sim t^{-3}.
\end{equation}

Let us consider a general power-law behavior of the CCDF, that is,
\begin{equation}
    \mathrm{Pr}(|X| > t) \sim t^{-\alpha},
\end{equation}
with $\alpha>0$ and for large $t$, i.e. $t > t_0$ for some tail onset scale $t_0$.
From the CCDF definition $\mathrm{Pr}(|X| > t) = \int_t^\infty f(x) dx$, it follows that the associated tail distribution  $f(t)$ scales as $f(t) = -(d/dt) \mathrm{Pr}(|X| > t) \sim t^{-\alpha - 1}$, so the tail contribution to the $n$-th moment behaves as
\begin{equation}
    \int_{t_0}^\infty x^n f(x) dx \sim \int_{t_0}^\infty x^{n-\alpha-1} dx. 
\end{equation}
This integral converges only if the upper limit $x^{n-\alpha}$ vanishes as $x \to \infty$.
This requires the exponent $n - \alpha$ to be negative, i.e. $n - \alpha < 0$ and so $\alpha > n$.
In particular, the mean ($n=1$) exists if $\alpha > 1$ and the second moment ($n=2$) if $\alpha > 2$.
Accordingly, this moment criterion ($\alpha > n$) is consistent with the variance scalings derived above and summarized in \cref{tab:toy_tail_exponents}.
In particular, for the second moment ($n=2$), $\alpha=3/2$ implies divergent variance (naive), $\alpha=2$ is the limit case leading to a logarithmic divergence (acceptance-weighted), and $\alpha=3$ implies finite variance (proposal-averaged, $K\to\infty$).
For finite proposal averaging (i.e. $2 \le K < \infty$), the effective tail exponent is expected to lie between the acceptance and infinite-$K$ scaling limits, i.e. $\alpha \in (2,3)$.
In that case, the $t^{-2}$ acceptance-like scaling still applies, but its prefactor is reduced by $1/K$, so increasing $K$ shifts the effective tail exponent toward the $t^{-3}$ scaling limit.

\begin{table}[h]
    \centering
    \begin{tabular}{cccc}
    \hline
    Estimator & $\alpha$ & Mean & Variance \\ \hline
    Naive & $3/2$ & finite & divergent \\
    Acceptance ($K=1$) & $2$ & finite & log-divergent \\
    Prop.-avg. ($K\to\infty$) & $3$ & finite & finite \\
    \hline
    \end{tabular}
    \caption{
    Asymptotic CCDF exponents ($\alpha$) for Pulay-type observables, defined from $\Pr(|X|>t)\sim t^{-\alpha}$ as $t\to\infty$, and the corresponding behavior of the first two moments.
    }
    \label{tab:toy_tail_exponents}
\end{table}

As an example, the results in \cref{fig:toy_model_force_distribution_and_ccdf} on the toy model described in \cref{app:toy_model_example} illustrate the expected hierarchy of tail softening, with the naive estimator displaying the heaviest tail, the acceptance-weighted estimator $(K=1)$ an intermediate behavior, and proposal averaging $(K=4,16,64,256)$ yielding the fastest decay.
The analysis focuses exclusively on the right tail, as the left tail decays exponentially for $\theta=0.25$, and therefore does not exhibit heavy-tailed (power-law) behavior.

The fitted exponents $\alpha$ reported in \cref{fig:toy_model_force_distribution_and_ccdf} are close to the analytical values in \cref{tab:toy_tail_exponents}.
They were obtained with TREAT \cite{2019_lopez_treat_code_github}, a tail-analysis code that estimates power-law exponents from the tails of a sampled distribution.
For each estimator, the analysis identifies a lower tail threshold and fits only the data above it.
The largest outliers are still included, so the threshold sets the beginning of the fitted tail region without imposing any upper cutoff.

These fitted exponents should be interpreted as finite-sample diagnostics of the analytical near-node scalings.
Because the sampled distributions are finite, the tail fit involves a bias-variance tradeoff: if the lower tail threshold is too low, the fit still includes data from the body of the distribution and has not yet reached the asymptotic regime; if it is too high, too few tail samples remain and the estimate becomes noisy.
In TREAT, the user typically does not set this threshold directly, but instead specifies an anchor search range from which the analysis selects it.
We therefore verified that the reported exponents remain stable under reasonable changes of the anchor range and systematically approach the analytical values as the number of Monte Carlo samples is increased.

\subsection{Time-dependent variational Monte Carlo example}

As a final example in this appendix, we turn to tVMC, where Pulay-like estimators with formally infinite variance naturally arise in the equations of motion \cite{2012_Carleo,2017_Carleo,2020_Schmitt,2021_Verdel,2023_Sinibaldi,2024_Nys,2025_Gravina}.
We follow the benchmark of Appendix E of Ref.~\cite{2017_Carleo}: a quench of $N=3$ particles at unit density, $N/L=1$, on a one-dimensional periodic lattice with $L/a=40$ sites, where $a$ is the lattice spacing, from the non-interacting ground state ($g=0$) to a final on-site interaction strength $g=8$.
This benchmark corresponds to a discretized form of the one-dimensional Lieb-Liniger bosonic model \cite{1963_lieb_liniger_bose_gas}.
The many-body wave function $\Phi(l_1,l_2,\dots,l_N,t)$, with each coordinate $l_i \in \boldsymbol{l} \equiv \{0,a,\dots,L-a\}$, evolves under the Hamiltonian
\begin{align}
    H_{a} \Phi(\boldsymbol{l}) &= -\frac{\hbar^2}{2ma^2} \sum_{i} \left\{ \frac{4}{3} \Big[\Phi(l_{i}-a) + \Phi(l_{i}+a) \Big]
    \right. \nonumber \\
    & \left. -\frac{1}{12} \Big[\Phi(l_{i}-2a) + \Phi(l_{i}+2a) \Big] - \frac{5}{2} \Phi(\boldsymbol{l}) \right\} \nonumber \\
    & + \frac{g}{a} \Phi(\boldsymbol{l}) \sum_{i<j} \delta(l_i,l_j)
\end{align}
where $\Phi(l_i-a) \equiv \Phi(l_1, \dots, l_i-a, \dots, l_N)$ (and similarly for $\Phi$ evaluated at the other shifted configurations) and $\delta(l_i,l_j)$ is the Kronecker delta.
The expressions in curly brackets correspond to the standard fourth-order central-difference approximation to the kinetic term, while the last term represents the discretized two-body contact interaction.

\begin{figure}[h]
  \centering
   \includegraphics[width=0.5\textwidth]{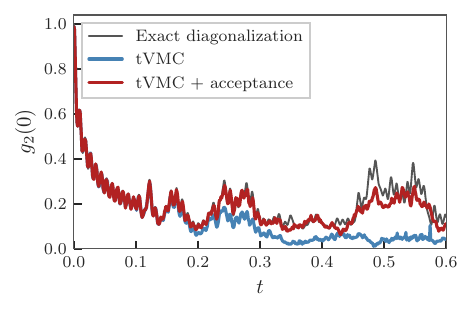}
  \caption{
  Time evolution of the double occupancy $g_2(r=0)$ for a quench of $N=3$ particles on a one-dimensional lattice with $40$ sites, from the non-interacting state at $t=0$ to an on-site interaction strength $g=8$.
  The curves compare bare tVMC, tVMC using the acceptance trick [\cref{eq:acceptance_trick_estimator}] in the equations of motion, and the exact diagonalization benchmark.}
  \label{fig:tVMC}
\end{figure}

We focus on the time-dependent double occupancy, $g_2(0) \equiv g_2(r=0,t)$, i.e. the pair-correlation function at zero distance.
This quantity is defined as $g_2(0) = 2L/(N(N-1)) \langle \sum_{i<j} \delta(l_i, l_j) / a \rangle$, corresponding to the contact operator normalized so that the initial non-interacting (delocalized) state satisfies $g_2(r=0,t=0)=1$.
A simple two-body Jastrow correlation yields a stable long-time evolution, but its limited expressivity introduces a noticeable bias.
More flexible parameterizations, such as generalized three-body correlations \cite{2017_Carleo} or neural-network states \cite{2017_carleo_nqs,2020_Schmitt,2021_Verdel,2023_Sinibaldi,2024_Nys,2025_Gravina}, substantially reduce this bias, but achieving long-time stability becomes much more challenging.
Because the matrices entering the tVMC equations of motion are Pulay-like estimators, outliers in their distributions can trigger instabilities in the dynamics.
Variance-reduction methods are therefore crucial \cite{2026_wan_blurred_sampling}.
Figure~\ref{fig:tVMC} shows the resulting improvement from applying the acceptance trick [\cref{eq:acceptance_trick_estimator}] in the time evolution of $g_2(0)$.
Following Ref.~\cite{2017_Carleo}, we integrate the equations of motion with a fourth-order Runge--Kutta method using a time step $dt=10^{-3}$.

\twocolumngrid
\bibliography{main.bib}

\clearpage

\onecolumngrid

\appendix

\section*{Supplemental Material}

\section{Near-node scaling of regularized Pulay estimators}
\label{app:near-node-scaling}

In this appendix, we continue the discussion started in Appendix A in the main manuscript, in particular, we analyze the near-node scaling of regularized Pulay-type observables.
We will re-use the same notation and assumptions from the first subsection of Appendix A.

\subsection{Hard-cutoff regularizations of the acceptance-weighted estimator}

Coming back to the acceptance-weighted estimator (though also relevant for the finite-$K$ proposal-averaged estimator, which has the same variance scaling), a simple way to make the variance finite for any fixed $\epsilon > 0$ is to multiply it by a cutoff that removes a neighborhood of the nodal surface.
Two natural choices are
\begin{equation} \label{app:hard_cutoff_regularization}
    \widetilde{O}_\epsilon^{(1)}(\mathbf{r}_i, \mathbf{r}_f) = \theta_\epsilon(\xi_i) \widetilde{O}(\mathbf{r}_i,\mathbf{r}_f),
\end{equation}
and
\begin{equation}
    \widetilde{O}_\epsilon^{(2)}(\mathbf{r}_i, \mathbf{r}_f) = \Theta_\epsilon(\xi_i, \xi_f) \widetilde{O}(\mathbf{r}_i, \mathbf{r}_f),
\end{equation}
where
\begin{equation}
    \theta_\epsilon(\xi) =
    \begin{cases}
    0, & 0 \le \xi < \epsilon, \\
    1, & \xi \ge \epsilon,
    \end{cases}
    \qquad
    \Theta_\epsilon(\xi_i, \xi_f) =
    \begin{cases}
    0, & 0 \le \xi_i < \epsilon \ \text{and}\ 0 \le \xi_f < \epsilon,\\
    1, & \text{otherwise},
    \end{cases}
\end{equation}
or equivalently, using the step function $\theta_\epsilon(\xi)$ to define $\Theta_\epsilon$ gives $\Theta_\epsilon(\xi_i, \xi_f) = 1 - [1-\theta_\epsilon(\xi_i)][1-\theta_\epsilon(\xi_f)]$.
The first excludes all samples whose current point lies within distance $\epsilon$ of the node.
The second is more selective and excludes only proposals for which both the current and proposed points lie in the ``nodal square'' $\{0 < \xi_i < \epsilon\} \cup \{0 < \xi_f < \epsilon\}$.

\subsubsection{Variance}

For $\widetilde{O}_\epsilon^{(1)}$, the lower bound of the $\xi_i$ integral is simply shifted from $\xi_c$ to $\epsilon$, so
\begin{equation}
    \boxed{
    \mathrm{Var} \left(\widetilde{O}_\epsilon^{(1)}\right) \sim \log \left( \frac{\xi_0}{\epsilon} \right)
    }
    \ .
\end{equation}
For $\widetilde{O}_\epsilon^{(2)}$, the square $0 < \xi_i < \epsilon$, $0 < \xi_f < \epsilon$ is excluded.
In region (A), the lower bound of the $\xi_f$-integral becomes $\max(\xi_i, \epsilon)$, so
\begin{equation}
    \langle (\widetilde{O}_\epsilon^{(2)})^2 \rangle_A  \sim \int_{0}^{\xi_0} d\xi_i \xi_i^2 \int_{\max(\xi_i,\epsilon)}^{\xi_0} d\xi_f \xi_f^{-4}.
\end{equation}
Splitting the outer integral into $0 < \xi_i < \epsilon$ and $\epsilon < \xi_i < \xi_0$, one finds that the former contributes only a finite constant, whereas the latter gives the same logarithm as before.
In region (B), the inequality $0 < \xi_f < \xi_i$ holds.
Therefore, if $\xi_i < \epsilon$, then necessarily $\xi_f < \epsilon$ as well, so the pair $(\xi_i,\xi_f)$ lies inside the square removed by $\Theta_\epsilon$.
Hence the condition $\Theta_\epsilon = 1$ forces $\xi_i > \epsilon$, and the contribution from region (B) starts at $\epsilon$, yielding once again $\int_\epsilon^{\xi_0} d\xi_i \xi_i^{-1}$. 
Hence
\begin{equation}
    \boxed{
    \mathrm{Var} \left( \widetilde{O}_\epsilon^{(2)} \right) \sim \log \left(\frac{\xi_0}{\epsilon} \right)
    }
    \ .
\end{equation}

\subsubsection{Bias}

For a regularized observable $O_\epsilon$, targeting the associated naive observable $O$, we define its bias with respect to the exact expectation value $\langle O \rangle$ as
\begin{equation}
    \mathrm{Bias}(O_\epsilon) \equiv \langle O_\epsilon \rangle - \langle O \rangle.
\end{equation}
Since the (unregularized) acceptance estimator is unbiased, the bias of the hard-cutoff estimators is simply the contribution removed by the cutoff.

For $\widetilde{O}_\epsilon^{(1)}$, the contribution removed comes from $0 < \xi_i < \epsilon$:
\begin{equation} \label{eq:bias_reg_acc_est_v1}
    \mathrm{Bias} \left(\widetilde{O}_\epsilon^{(1)} \right) \sim \int_0^\epsilon d\xi_i \xi_i^2 \left[\int_{\xi_i}^{\xi_0} d\xi_f \xi_f^{-2} + \int_0^{\xi_i} d\xi_f \xi_i^{-2} \right].
\end{equation}
The first term corresponds to region (A), where $a=1$ and $\widetilde{O} \sim \xi_f^{-2}$, while the second corresponds to region (B), where $\widetilde{O} \sim \xi_i^{-2}$. 
Since $\int_{\xi_i}^{\xi_0} d\xi_f \xi_f^{-2} \sim \xi_i^{-1}$ and $\int_0^{\xi_i} d\xi_f \xi_i^{-2} \sim \xi_i^{-1}$, one finds
\begin{equation}
    \boxed{
    \mathrm{Bias} \left(\widetilde{O}_\epsilon^{(1)} \right) = \mathcal{O}(\epsilon^2)
    }
    \ .
\end{equation}
For the more selective cutoff $\widetilde{O}_\epsilon^{(2)}$, the removed contribution is restricted to the square $0 < \xi_i < \epsilon$ and $0 < \xi_f < \epsilon$, so that
\begin{equation}
    \mathrm{Bias} \left(\widetilde{O}_\epsilon^{(2)} \right) \sim \int_0^\epsilon d\xi_i \xi_i^2 \left[\int_{\xi_i}^{\epsilon} d\xi_f \xi_f^{-2} + \int_0^{\xi_i} d\xi_f \xi_i^{-2} \right],
\end{equation}
where the first and second terms correspond to regions (A) and (B), respectively.
Therefore, both contributions scale as $\int_0^\epsilon d\xi_i \xi_i$, yielding
\begin{equation}
    \boxed{
    \mathrm{Bias} \left(\widetilde{O}_\epsilon^{(2)} \right) = \mathcal{O}(\epsilon^2)
    }
    \ ,
\end{equation}
though with a smaller prefactor as compared to $\widetilde{O}_\epsilon^{(1)}$, so $\mathrm{Bias}(\widetilde{O}_\epsilon^{(2)}) < \mathrm{Bias}(\widetilde{O}_\epsilon^{(1)})$.
This is obviously because $\widetilde{O}_\epsilon^{(1)}$ discards all points with $0 < \xi_i < \epsilon$, independently of $\xi_f$, excluding the entire strip $\{0 < \xi_i < \epsilon \} \cup \{ 0 < \xi_f < \xi_0 \}$, while $\widetilde{O}_\epsilon^{(2)}$ discards ``only'' the square $\{0 < \xi_i < \epsilon\} \cup \{0 < \xi_f < \epsilon\}$.

\subsection{Smooth cutoffs and systematic cancellation of the bias}

The quadratic bias of a hard cutoff is not fundamental. 
It arises because the leading contribution removed from the nodal strip is linear in the distance to the node (i.e. $\mathrm{Bias} \sim \int_0^\epsilon \xi d\xi \sim \epsilon^2$).
This can be changed by replacing the sharp cutoff by a smooth function \cite{2020_pathak_wagner_gradient_regularization}.
We consider
\begin{equation} \label{eq:smooth_cutoff_pulay_acceptance_estimator}
    \widehat{O}_\epsilon(\mathbf{r}_i, \mathbf{r}_f) = \chi_\epsilon(\xi_i) \widetilde{O}(\mathbf{r}_i, \mathbf{r}_f), \qquad \text{where} \qquad 
    \chi_\epsilon(\xi) =
    \begin{cases}
    \chi(\xi/\epsilon), & 0 \le \xi < \epsilon, \\
    1, & \xi \ge \epsilon,
    \end{cases}
\end{equation}
for some function $\chi: [0,1] \to \mathbb R$ such that $\chi(0) = 0$ and $\chi(1) = 1$.
We may additionally impose smooth matching conditions such as $\chi'(0) = 0$ and $\chi'(1) = 0$.
(The generalization to a two-point smooth cutoff estimator $\widehat{O}_\epsilon^{(2)}(\mathbf{r}_i, \mathbf{r}_f) = \varphi_\epsilon(\xi_i, \xi_f) \widetilde{O}(\mathbf{r}_i, \mathbf{r}_f)$, for some smooth cutoff function $\varphi_\epsilon$, will not be pursued here.)

To analyze the bias, we define the following contribution to the mean of the acceptance estimator by
\begin{equation}
    G(\xi_i) \equiv \int d\mathbf{r}_{i,\|} \int d\mathbf{r}_f |\Psi(\mathbf{r}_i)|^2 p(\mathbf{r}_f|\mathbf{r}_i) \widetilde{O}(\mathbf{r}_i, \mathbf{r}_f),
\end{equation}
where $\mathbf{r}_i = (\xi_i, \mathbf{r}_{i,\|})$ and $\xi_i > 0$.
In other words, $G(\xi_i)$ is the contribution to the expectation value of the acceptance-weighted estimator after integrating out all tangential coordinates and the proposed configuration $\mathbf{r}_f$, while keeping fixed the normal distance $\xi_i$ of the current configuration to the nodal surface.

We saw for the hard-cutoff bias analysis that the contribution removed from the strip scales like $\int_0^\epsilon d\xi_i G(\xi_i) \sim \epsilon^2$.
For that to happen, $G(\xi_i)$ itself must scale linearly at leading order, that is, $G(\xi_i) \sim \xi_i$.
Assuming $G$ is smooth enough near $\xi_i = 0$, its Taylor expansion is
\begin{equation}
    G(\xi_i) = c_1 \xi_i + c_2 \xi_i^2 + \mathcal{O}(\xi_i^3) = \sum_{n \ge 1} c_n \xi_i^n.
\end{equation}
The analogous expression of the $\widetilde{O}_\epsilon^{(1)}$-bias given in \cref{eq:bias_reg_acc_est_v1} for the smooth-cutoff estimator $\widehat{O}_\epsilon$ is then
\begin{equation} 
    \mathrm{Bias}(\widehat{O}_\epsilon) = \int_0^\epsilon d\xi [\chi(\xi/\epsilon) - 1] G(\xi).
\end{equation}
Introducing $t = \xi/\epsilon$, so $d\xi = \epsilon dt$, we obtain
\begin{align} \label{eq:bias_expansion_smooth_cutoff_estimator}
    \mathrm{Bias}(\widehat{O}_\epsilon) 
    &= \int_0^\epsilon d\xi \big[ \chi(\xi/\epsilon) - 1 \big] \big[c_1 \xi + c_2 \xi^2 + \mathcal{O}(\xi^3) \big] \nonumber\\
    &= \sum_{n \ge 1} c_n \epsilon^{n+1} \int_0^1 dt [\chi(t) - 1] t^n \nonumber\\
    &= c_1\epsilon^2 \int_0^1 dt [\chi(t) - 1] t + c_2\epsilon^3 \int_0^1 dt [\chi(t) - 1] t^2 + \mathcal{O}(\epsilon^4).
\end{align}
Therefore, if $\chi$ is chosen such that $\int_0^1 dt [\chi(t) - 1] t = 0$, the quadratic term $\sim \epsilon^2$ vanishes and the leading bias becomes cubic.
A simple quartic example is
\begin{equation} \label{eq:quartic_polynomial_canceling_quadratic_term}
    \chi(t) = 12t^2 - 20t^3 + 9t^4, \qquad 0 \le t \le 1,
\end{equation}
which satisfies $\chi(0) = 0$, $\chi'(0)=0$, $\chi(1) = 1$, $\chi'(1)=0$, and $\int_0^1 dt [\chi(t) - 1] t = 0$.
Its bias then scales as 
\begin{equation}
    \boxed{
    \mathrm{Bias}(\widehat{O}_\epsilon) = \mathcal{O}(\epsilon^3)
    }
    \ .
\end{equation}
Because $\chi_\epsilon(\xi)$ still vanishes at the node and equals $1$ outside the $\epsilon$-neighborhood, the inner region contributes only a finite constant to the second moment, while the outer region retains the same logarithmic behavior as before, so
\begin{equation}
    \boxed{
    \mathrm{Var}(\widehat{O}_\epsilon) \sim \log \left(\frac{\xi_0}{\epsilon} \right).
    }
\end{equation}

\subsection{Polynomial ansatz and higher-order bias cancellation}

While the smooth cutoff need not be polynomial (it could be a spline, a trigonometric function or another smooth function), a polynomial ansatz is convenient because all constraints reduce to linear equations for its coefficients so that the bias-cancellation conditions can be imposed analytically. 
Let
\begin{equation} \label{eq:cutoff_polynomial_ansatz}
    \chi(t) = \sum_{j=0}^{d} a_j t^j, \qquad 0 \le t\le 1.
\end{equation}
If we want the bias to start at order $\epsilon^{m+2}$ ($m = 1, 2, \dots$), then the first $m$ moments must vanish:
\begin{equation} \label{eq:cutoff_moment_conditions}
    M_n[\chi] \equiv \int_0^1 dt [\chi(t) - 1] t^n = 0, \qquad n=1,\dots,m.
\end{equation}
Together with endpoint conditions such as
\begin{equation}
    \chi(0) = 0, \qquad \chi'(0)=0, \qquad \chi(1) = 1, \qquad \chi'(1) = 0,
\end{equation}
these constraints define a linear system for the coefficients $\{a_j\}$.
The endpoint conditions give, for example,
\begin{equation}
    a_0=0, \qquad a_1=0, \qquad \sum_{j=0}^{d} a_j = 1, \qquad \sum_{j=1}^{d} j a_j = 0.
\end{equation}
Likewise, substituting the polynomial ansatz \cref{eq:cutoff_polynomial_ansatz} into the moment condition \cref{eq:cutoff_moment_conditions} yields
\begin{equation}
    \int_0^1 dt \chi(t) t^n = \sum_{j=0}^{d} a_j \int_0^1 dt t^{n+j} = \sum_{j=0}^{d} \frac{a_j}{n+j+1}.
\end{equation}
Since
\begin{equation}
    M_n[\chi] = 0 \iff \int_0^1 dt \chi(t) t^n = \int_0^1 t^n dt = \frac{1}{n+1},
\end{equation}
each moment constraint becomes the linear equation
\begin{equation}
    \sum_{j=0}^{d}\frac{a_j}{j+n+1} = \frac{1}{n+1}, \qquad n=1,\dots,m.
\end{equation}
Collecting all endpoint and moment conditions gives a matrix equation
\begin{equation}
    A \mathbf{a} = \mathbf{b},
\end{equation}
with $\mathbf{a} = (a_0,\dots,a_d)^T$ and $\mathbf{b}$ the vector encoding the constraints.
If $d$ is larger than strictly necessary, that is, $d + 1 > N_c$, where $N_c = m + 4$ is the total number of constraints, the system is underdetermined and one obtains a family of admissible smooth cutoffs.
For the example given in \cref{eq:quartic_polynomial_canceling_quadratic_term}, the matrix $A$ is square and has a unique solution since a degree-4 polynomial has 5 coefficients and we imposed 5 constraints.
In practice, canceling only the leading moment is often already sufficient and provides a good compromise between bias reduction and numerical robustness (as the smooth cutoff becomes increasingly oscillatory (in the interval $0 \le t < 1$) as more moments are canceled).

\subsection{Practical choice of the regularization parameter $\epsilon$} \label{app:practical_choice_of_epsilon}

In practice, the cutoff parameter $\epsilon$ plays the role of a bias-variance trade-off parameter.
As shown above, the smooth-cutoff regularized estimators satisfy
\begin{equation}
    \mathrm{Var}(\widehat{O}_\epsilon) \sim \log(\xi_0/\epsilon), \qquad \mathrm{Bias}(\widehat{O}_\epsilon) \sim \epsilon^{m+2},
\end{equation}
where, again, $m$ is the number of canceled bias moments of the cutoff function.
The mean-squared error (MSE) therefore behaves as 
\begin{equation}
    \mathrm{MSE}(\epsilon) = \underbrace{\mathrm{Var}(\widehat{O}_\epsilon)/N}_{\text{statistical error}^2} + \underbrace{\mathrm{Bias}(\widehat{O}_\epsilon)^2}_{\text{systematic error}^2} \sim \log(\xi_0/\epsilon) / N + \epsilon^{2(m+2)},
\end{equation}
which shows that decreasing $\epsilon$ reduces the bias but increases the variance.

In the limit of $\epsilon \to 0$, the estimator approaches the unregularized one and the variance reduction disappears.
Conversely, choosing $\epsilon$ too large suppresses the outliers but introduces a sizeable systematic bias.
A practical value of $\epsilon$ must therefore lie in an intermediate regime where the regularization is ``active'' but the bias remains negligible compared to the statistical uncertainty.
In practice, $\epsilon$ can be chosen by evaluating the estimator for a small set of (decreasing) $\epsilon$ values.
The optimal region is identified when the mean stabilizes, while the variance remains reduced compared to the unregularized estimator.
One then selects $\epsilon$ as the largest value for which the bias is below the statistical error, while still providing some variance reduction and thus a non-zero fraction of regularized samples.

\section{Gradient-to-Laplacian identity} \label{app:gradient_to_laplacian_identity}

We derive the gradient-to-Laplacian identity used in the main manuscript to obtain improved Hellmann-Feynman force estimators, that we rewrite here for completeness,
\begin{equation} \label{eq:gradient_to_laplacian_identity_app}
    2 \partial_{x_\alpha} v(x) = \nabla_{\mathbf{x}}^2 \big[x_\alpha v(x) \big] - \frac{x_\alpha}{x} f''(x), \qquad f(x)\equiv x v(x),
\end{equation}
valid for $x\neq 0$ in $\mathbb R^3$, with $x=|\mathbf x|$ and for some function $v(x)$.
Using $\partial_{x_\beta}x_\alpha=\delta_{\alpha\beta}$ and Einstein summation notation, we get
\begin{align} \label{eq:lap_x_v_app}
    \nabla_{\mathbf x}^2 \big[x_\alpha v(x)\big]
    &= \partial_{x_\beta} \partial_{x_\beta} \big[x_\alpha v(x)\big] \nonumber \\
    &= \partial_{x_\beta} \left(\delta_{\alpha\beta} v(x) + x_\alpha \partial_{x_\beta} v(x) \right) \nonumber\\
    &= \partial_{x_\alpha} v(x) + \partial_{x_\beta} \big[x_\alpha \partial_{x_\beta} v(x) \big] \nonumber \\
    &= \partial_{x_\alpha} v(x) + \delta_{\alpha\beta} \partial_{x_\beta} v(x) + x_\alpha \partial_{x_\beta} \partial_{x_\beta} v(x) \nonumber\\
    &= 2 \partial_{x_\alpha} v(x) + x_\alpha\nabla_{\mathbf x}^2 v(x). 
\end{align}
The radial Laplacian (in three-dimension) is derived as follows
\begin{align}
    \nabla_{\mathbf{x}} v(x) &= v'(x) \nabla_{\mathbf{x}} x = v'(x) \frac{\mathbf{x}}{x}, \\
    \nabla_{\mathbf{x}}^2 v(x) &= \nabla_{\mathbf{x}} \cdot \left(v'(x) \frac{\mathbf{x}}{x}\right) = \nabla_{\mathbf{x}} v'(x) \cdot\frac{\mathbf{x}}{x} + v'(x) \nabla_{\mathbf{x}} \cdot \left(\frac{\mathbf{x}}{x}\right) = v''(x) + \frac{2}{x}v'(x) \label{eq:radial_lap_app},
\end{align}
where we used $\nabla_{\mathbf{x}} v'(x) = v''(x) \mathbf{x} / x$ and $\nabla_{\mathbf{x}} \cdot (\mathbf{x}/x) = (\nabla_\mathbf{x} \cdot \mathbf{x})/x + \mathbf{x} \cdot \nabla_\mathbf{x}(1/x) = 3/x - 1/x = 2/x$ (since $\nabla_\mathbf{x} (1/x) = -\mathbf{x}/x^3$).
Now introduce $f(x)=xv(x)$. A direct differentiation gives
\begin{equation}
    f''(x) = \frac{d^2}{dx^2} \big[ xv(x) \big] = \frac{d}{dx}\big[ v(x) + x v'(x) \big] = 2 v'(x) + x v''(x),
\end{equation}
so \cref{eq:radial_lap_app} can be written as
\begin{equation} \label{eq:lap_in_terms_of_f_app}
    \nabla_{\mathbf{x}}^2 v(x) = \frac{1}{x} f''(x).
\end{equation}
Substituting \cref{eq:lap_in_terms_of_f_app} into \cref{eq:lap_x_v_app} yields
\begin{equation}
    \nabla_{\mathbf{x}}^2 \big[ x_\alpha v(x) \big] = 2 \partial_{x_\alpha} v(x) + \frac{x_\alpha}{x} f''(x),
\end{equation}
which rearranges to \cref{eq:gradient_to_laplacian_identity_app}.

\section{Divergence theorem identities} \label{app:divergence_theorem_identities}

We adopt in this appendix a more math-based notation, in particular without putting vectors in bold font.
Let $x \in \Omega$ denote a configuration-space sample (e.g. $x \equiv \mathbf{r} \in \mathbb{R}^{3N}$ in VMC), with domain $\Omega = \mathbb{R}^d$ for open boundary conditions, or a periodic simulation cell under periodic boundary conditions.
Let $\pi(x) \ge 0$ be an unnormalized probability distribution on $\Omega$, with normalization $Z = \int_\Omega \pi(x) dx$. 
For an observable $O$, we write $\langle O \rangle \equiv \langle O \rangle_\pi = Z^{-1} \int_\Omega O(x) \pi(x) dx$. 
For a sufficiently regular vector field $A : \Omega \to \mathbb{R}^d$, define the Langevin Stein operator \cite{2022_si_control_variate_for_mc_methods}
\begin{equation}
    \mathcal{L}_\pi[A](x) \equiv \nabla \cdot A(x) + A(x) \cdot \nabla \log \pi(x),
\end{equation}
where $\nabla \equiv \nabla_x$ is the configuration-space gradient. 
Throughout this section, we assume boundary conditions such that the divergence theorem can be applied to the vector field entering each identity without producing residual boundary terms.
Concretely, for open boundary conditions, it means that the relevant field $\pi(x) A(x) H(x)$, defined below for some indicator function $H$, decays sufficiently rapidly as $|x| \to \infty$, while for periodic boundary conditions, $\pi(x), A(x)$ and $H(x)$ are periodic on $\Omega$.
In each subsection below, we present an identity with an associated (boxed) example.

\subsection{Codimension-0: zero-mean control variate} \label{sec:codimension_0_ibp_identity}

Consider the vector field 
\begin{equation}
    F(x) = \pi(x) A(x).
\end{equation}
By the divergence theorem and the above boundary assumptions,
\begin{equation}
    \int_\Omega \nabla \cdot F(x) dx = \int_{\partial \Omega} F \cdot n dS = 0.
\end{equation}
Using $\nabla \cdot (\pi A) = \pi(\nabla \cdot A + A \cdot \nabla \log \pi) = \pi \mathcal{L}_\pi[A]$, we obtain $\langle \mathcal{L}_\pi[A] \rangle = 0$, so for any observable $O$,
\begin{equation} \label{eq:codimension_0_idendity}
    \boxed{
    \langle O \rangle = \langle O + \mathcal{L}_\pi[A] \rangle
    }
    \ .
\end{equation}
Thus $\mathcal{L}_\pi[A]$ is a zero-mean control variate correction, which can reduce the estimator variance when the vector field $A$ is chosen appropriately. \\

\begin{mybox}{Example: forces under open boundary conditions}{codimension_0_example_forces_obc}
Consider $x \equiv \mathbf{r} = (\mathbf{r}_1, \hdots, \mathbf{r}_N)$, $\pi(\mathbf{r}) = |\Psi(\mathbf{r})|^2$, and a nucleus $I$ of charge $Z_I$ at a fixed position $\mathbf{R}_I$.
We denote the gradient and Laplacian respectively as $\nabla \equiv \nabla_\mathbf{r}$ and $\nabla^2 \equiv \sum_{i\alpha} \partial^2 / \partial_{r_{i \alpha}^2}$.
The bare electron-nucleus Hellmann-Feynman force on nucleus $I$ along the direction $\alpha$ is 
\begin{equation} \label{eq:HF_force_app}
    \vspace{-0.4em}
    F_{I \alpha, en}^\mathrm{HF}(\mathbf{r}) = Z_I \sum_i \frac{r_{i\alpha} - R_{I\alpha}}{|\mathbf{r}_i - \mathbf{R}_I|^3},
    \vspace{-0.4em}
\end{equation}
which is singular as $|\mathbf{r}_i - \mathbf{R}_I| \to 0$.
The choice of $\mathbf{A}$ made in the work of Assaraf and Caffarel \cite{2003_assaraf_caffarel_zero_variance_force} is
\begin{equation}
    \vspace{-0.4em}
    \mathbf{A}_{I \alpha}(\mathbf{r}) = \nabla Q_{I \alpha}(\mathbf{r}), \qquad Q_{I\alpha}(\mathbf{r}) = Z_I \sum_i \frac{r_{i\alpha} - R_{I\alpha}}{|\mathbf{r}_i - \mathbf{R}_I|}.
    \vspace{-0.4em}
\end{equation}
Note that with a reasonable wave function choice, $\pi(\mathbf{r}) \mathbf{A}_{I\alpha}(\mathbf{r})$ decays sufficiently fast at infinity so that the boundary term vanishes, thus satisfying the boundary assumption.
Using the identity $\nabla^2((r_{i\alpha} - R_{I\alpha})/|\mathbf{r}_i - \mathbf{R}_I|) = -2(r_{i\alpha} - R_{I\alpha})/|\mathbf{r}_i - \mathbf{R}_I|^3$ (for $\mathbf{r}_i \ne \mathbf{R}_I$), one then finds that $\nabla \cdot \mathbf{A}_{I\alpha} = \nabla^2 Q_{I \alpha} = -2 F_{I\alpha,en}^\mathrm{HF}$, meaning that this divergence part of $\mathcal{L}_\pi[\mathbf{A}_{I\alpha}]$ cancels the Coulomb singular term in \cref{eq:HF_force_app} exactly (provided it is rescaled by a factor of 2, which is possible because by construction $\langle \mathcal{L}_\pi[\mathbf{A}_{I\alpha}] \rangle = 0$ implies that $\langle \mathcal{L}_\pi[\mathbf{A}_{I\alpha}]/2 \rangle = 0$). 
Inserting into \cref{eq:codimension_0_idendity}, the corresponding improved estimator is then 
\begin{equation}
    F_{I\alpha}^\mathrm{imp}(\mathbf{r}) 
    = F_{I\alpha}^\mathrm{HF}(\mathbf{r}) + \mathcal{L}_\pi[\mathbf{A}_{I\alpha}](\mathbf{r})/2
    = F_{I\alpha,nn}^\mathrm{HF}(\mathbf{R}) + \nabla Q_{I \alpha}(\mathbf{r}) \cdot \nabla \log |\Psi(\mathbf{r})|,
\end{equation}
which is unbiased because $\langle \mathcal{L}_\pi[\mathbf{A}] \rangle = 0$, and it has much smaller variance because the singular term $F_{I \alpha, en}^\mathrm{HF}(\mathbf{r})$, given in \cref{eq:HF_force_app}, is exactly canceled, and only the constant nucleus-nucleus term $F_{I \alpha, nn}^\mathrm{HF}(\mathbf{R})$ survives from the bare estimator $F_{I\alpha}^\mathrm{HF}(\mathbf{r})$.
\end{mybox}
\vspace{0.2cm}

\subsection{Codimension-1: surface-to-volume conversion} \label{sec:codimension_1_ibp_identity}

Let $\xi: \Omega \to \mathbb{R}$ be differentiable, fix $s \in \mathbb{R}$, and define the Heaviside step function $\theta_s(x) \equiv \theta(s - \xi(x))$, which should be periodized accordingly under periodic boundary conditions.
Consider the truncated vector field
\begin{equation}
    F(x) = \pi(x) \theta_s(x) A(x).
\end{equation}
Applying the divergence theorem on $\Omega$ gives, as before, $\int_\Omega \nabla \cdot F dx = 0$.
Expanding $\nabla \cdot F = \nabla \cdot (\pi \theta_s A)$ via the product rule yields
\begin{equation}
    \nabla \cdot (\pi \theta_s A) = \pi \theta_s \mathcal{L}_\pi[A] + \pi (\nabla \theta_s) \cdot A.
\end{equation}
Using $\nabla \theta_s = \nabla_x \theta(s - \xi(x)) = -\delta(s - \xi) \nabla \xi$ (since $\partial_t \theta(t) = \delta(t)$), we obtain the identity
\begin{equation} \label{eq:codimension_1_identity}
    \boxed{
    \langle \delta(s - \xi) A \cdot \nabla \xi \rangle = \langle \theta_s \mathcal{L}_\pi[A] \rangle
    }
    \ .
\end{equation}
For observables involving a single delta function of the form $\langle \delta(s - \xi(x)) W(x) \rangle$ (e.g. (multipole) densities, pair correlations, angular distributions), one typically chooses $A$ such that on $\{ \xi = s \}$, $A(x;s) \cdot \nabla \xi(x) = W(x)$.
A simple choice is 
\begin{equation}
    \boxed{
    A(x;s) = W(x) \frac{\nabla \xi(x)}{|\nabla \xi(x)|^2}
    }
    \ ,
\end{equation}
provided $\nabla \xi \ne 0$.
With this choice, the left-hand side of \cref{eq:codimension_1_identity} reduces to $\langle \delta(s - \xi) W \rangle$, while the right-hand side is a sublevel-set (volume) average over $\{\xi \le s\}$.
We emphasize that \cref{eq:codimension_1_identity} turns an integration over a shell into an integration over a ball, therefore increasing the signal-to-noise ratio in numerical calculations. 
\\

\vspace{0.2cm}
\begin{mybox}{Example: pair correlation functions}{codimension_1_example_pair_correlation_function}
For a system of particles with positions $\{\mathbf{r}_i\}$, the bare radial pair correlation function estimator is $g(s) \propto \sum_{i<j} \delta(s - r_{ij})$, $r_{ij} = |\mathbf{r}_i - \mathbf{r}_j|$.
To apply \cref{eq:codimension_1_identity}, fix a pair $(i,j)$ and set $\xi(x) = r_{ij}(x)$. 
The configuration space gradient is 
\begin{equation}
    \nabla r_{ij} = (0, \hdots, \underbrace{\hat{\mathbf{r}}_{ij}}_{\mathbf{r}_i \ \text{block}}, \hdots, \underbrace{-\hat{\mathbf{r}}_{ij}}_{\mathbf{r}_j \ \text{block}}, \hdots, 0), \qquad \hat{\mathbf{r}}_{ij} = \frac{\mathbf{r}_i - \mathbf{r}_j}{r_{ij}},
\end{equation}
so that $|\nabla r_{ij}|^2  = |\hat{\mathbf{r}}_{ij}|^2 + |-\hat{\mathbf{r}}_{ij}|^2 = 2$.
A convenient choice of the vector field satisfying $\mathbf{A}_{ij} \cdot \nabla r_{ij} = 1$ on $\{r_{ij}=s\}$ is therefore
\begin{equation}
    \mathbf{A}_{ij}(x) = \frac{\nabla r_{ij}}{|\nabla r_{ij}|^2} = \frac{1}{2} (0, \hdots, \hat{\mathbf{r}}_{ij}, \hdots, -\hat{\mathbf{r}}_{ij}, \hdots, 0).
\end{equation}
Note that, to further regularize short-distances, one can in principle scale $A_{ij}$ by a smooth shaping factor $g(r_{ij}; s)$, provided $g(s;s)=1$.
We now evaluate $\mathcal{L}_\pi[\mathbf{A}_{ij}] = \nabla \cdot \mathbf{A}_{ij} + \mathbf{A}_{ij} \cdot \nabla \log \pi$.
Using $\nabla_{\mathbf{r}_i} \cdot \hat{\mathbf{r}}_{ij} = 2 / r_{ij}$ and $\nabla_{\mathbf{r}_j} \cdot \hat{\mathbf{r}}_{ij} = -2 / r_{ij}$ we obtain $\nabla \cdot \mathbf{A}_{ij} = 2 / r_{ij}$.
The second term becomes a relative drift term, so inserting these expressions into \cref{eq:codimension_1_identity} gives, for a single pair $(i,j)$,
\begin{equation}
    \langle \delta(s - r_{ij}) \rangle =  \left\langle \theta(s - r_{ij}) \left( \frac{2}{r_{ij}} + \hat{\mathbf{r}}_{ij} \cdot (\nabla_i - \nabla_j) \log \Psi \right) \right\rangle \qquad (\text{VMC, } \pi = |\Psi|^2).
\end{equation}
For classical NVT sampling methods, such as classical Monte Carlo (MC) or molecular dynamics (MD), with $\pi \propto e^{-\beta U}$ and $\nabla_i \log \pi = \beta \mathbf{F}_i^\mathrm{cl}$, the improved estimator is
\begin{equation}
    \langle \delta(s - r_{ij}) \rangle =  \left\langle \theta(s - r_{ij}) \left( \frac{2}{r_{ij}} + \frac{\beta}{2} \hat{\mathbf{r}}_{ij} \cdot (\mathbf{F}_i^\mathrm{cl} - \mathbf{F}_j^\mathrm{cl}) \right) \right\rangle \qquad (\text{MD/MC, } \pi = e^{-\beta U}).
\end{equation}
Summing over all pairs $(i<j)$ and applying the proper normalization yields the corresponding improved estimator for $g(s)$.
\end{mybox}
\vspace{0.2cm}

\subsection{Codimension-2: intersection-to-volume conversion}
\label{sec:codimension_2_ibp_identity}

We now consider observables involving two delta constraints of the generic form
\begin{equation} \label{eq:two_delta_function_estimator_form}
    g(s,t) = \big\langle \delta(s - \xi(x)) \delta(t - \eta(x)) \big\rangle,
\end{equation}
for differentiable function $\xi, \eta: \Omega \to \mathbb{R}$ and constants $s, t \in \mathbb{R}$.
These estimators are supported on the intersection $\{\xi = s\} \cap \{\eta = t\}$.
Define
\begin{equation}
    \theta_s(x) \equiv \theta(s - \xi(x)), \qquad \theta_t(x) \equiv \theta(t - \eta(x)), \qquad \delta_s(x) = \delta(s - \xi(x)), \qquad \delta_t(x) \equiv \delta(t - \eta(x)),
\end{equation}
and let $p = \nabla \xi$, $q = \nabla \eta$, so that $\nabla \theta_s = -\delta_s p$ and $\nabla \theta_t = -\delta_t q$. \\

\textbf{First elimination.}
Applying the divergence theorem to the doubly truncated vector field
\begin{equation}
    F_A(x) = \pi(x) \theta_s(x) \theta_t(x) A(x),
\end{equation}
yields $\int_\Omega \nabla \cdot F_A dx = 0$, given our initial boundary assumptions.
Expanding gives
\begin{equation}
    \nabla \cdot (\pi \theta_s \theta_t A) = \pi \theta_s \theta_t \mathcal{L}_\pi[A] - \pi \delta_s \theta_t (A \cdot p) - \pi \theta_s \delta_t (A \cdot q).
\end{equation}
To eliminate one surface term, without loss of generality, we choose $A$ such that (we will come back to this choice later, causing asymmetry in the final expression)
\begin{equation} \label{eq:codimension2_ibp_constraints_on_A}
    A \cdot p = 1 \qquad \text{and} \qquad A \cdot q =0.
\end{equation}
In that case, integrating over $\Omega$ and dividing by the normalization $Z$, we get 
\begin{equation}
    \langle \delta_s \theta_t \rangle = \langle \theta_s \theta_t \mathcal{L}_\pi[A] \rangle.
\end{equation}
We now differentiate this expression with respect to $t$, noting that $A$ and $\mathcal{L}_\pi[A]$ are independent of $t$, to reach the intermediate expression
\begin{equation} \label{eq:codimension_2_ibp_identity_inter_expr_1}
    \langle \delta_s \delta_t \rangle = \langle \theta_s \delta_t \mathcal{L}_\pi[A] \rangle.
\end{equation} \\

\textbf{Second elimination.}
As we are still left with one delta function on the right-hand side of \cref{eq:codimension_2_ibp_identity_inter_expr_1}, we proceed to eliminate it by applying a second time the divergence theorem, but this time to the new vector field (defined in such a way to be able to use the intermediate identity \cref{eq:codimension_2_ibp_identity_inter_expr_1})
\begin{equation}
    F_B(x) = \pi(x) \theta_s(x) \theta_t(x) B(x) \mathcal{L}_\pi[A](x).
\end{equation}
Applying the divergence theorem to $F_B$ gives $\int_\Omega \nabla \cdot F_B dx = 0$.
Expanding again using the product rule, we get two volume and two surface terms 
\begin{equation}
    \nabla \cdot (\pi \theta_s \theta_t B \mathcal{L}_\pi[A]) = \pi \theta_s \theta_t \Big\{ \mathcal{L}_\pi[B] \mathcal{L}_\pi[A] + B \cdot \nabla \mathcal{L}_\pi[A] \Big\} -\pi \Big\{ \delta_s \theta_t (B \cdot p) + \theta_s \delta_t (B \cdot q) \Big\} \mathcal{L}_\pi[A].
\end{equation}
To be able to use the identity given in \cref{eq:codimension_2_ibp_identity_inter_expr_1}, we choose $B$ such that
\begin{equation} \label{eq:codimension2_ibp_constraints_on_B}
    B \cdot p = 0 \qquad \text{and} \qquad B \cdot q = 1.
\end{equation}
Applying these constraints, integrating over $\Omega$ and dividing by $Z$, we get
\begin{equation} \label{eq:codimension_2_ibp_identity_inter_expr_2}
    \langle \theta_s \delta_t \mathcal{L}_\pi[A] \rangle = \langle \theta_s \theta_t (\mathcal{L}_\pi[B] \mathcal{L}_\pi[A] + B \cdot \nabla \mathcal{L}_\pi[A]) \rangle.
\end{equation}
We can finally combine \cref{eq:codimension_2_ibp_identity_inter_expr_1,eq:codimension_2_ibp_identity_inter_expr_2} to obtain an expression that is now free of delta functions, that is,
\begin{equation}
    \langle \delta_s \delta_t \rangle = \langle \theta_s \theta_t (\mathcal{L}_\pi[B] \mathcal{L}_\pi[A] + B \cdot \nabla \mathcal{L}_\pi[A]) \rangle.
\end{equation}
Given that we eliminated the two delta functions sequentially, the expression is manifestly asymmetric, as previously alluded to.
This can be mathematically seen even better by writing the expression in parentheses as $\mathcal{L}_\pi[B] \mathcal{L}_\pi[A] + B \cdot \nabla \mathcal{L}_\pi[A] = \mathcal{L}_\pi[B \mathcal{L}_\pi[A]]$, the latter clearly identifying that $B$ was introduced after weighting by $\mathcal{L}_\pi[A]$.
If instead $\delta_t$ is first removed with $B$ and then $\delta_s$ is removed with $A$, the two theta functions would be multiplied by $\mathcal{L}_\pi[A \mathcal{L}_\pi[B]]$ instead of $\mathcal{L}_\pi[B \mathcal{L}_\pi[A]]$.
We can therefore average the two orderings to obtain the symmetric formula
\begin{equation} \label{eq:codimension_2_identity}
    \boxed{
    \langle \delta_s \delta_t \rangle = \left\langle \theta_s \theta_t \frac{1}{2} (\mathcal{L}_\pi[A \mathcal{L}_\pi[B]] + \mathcal{L}_\pi[B \mathcal{L}_\pi[A]]) \right\rangle
    }
    \ ,
\end{equation}
which is now indeed manifestly symmetric under swapping $(A, \xi, s) \leftrightarrow (B, \eta, t)$.
In addition, one can use the identity $\mathcal{L}_\pi[\phi A] = \mathcal{L}_\pi[A] \phi + A \cdot \nabla \phi$, for some scalar $\phi$, to expand this expression (and facilitate computer implementation). 
\\

\textbf{Constructing the vector fields $A$ and $B$}.
We started with the level-set constraints $\xi(x) = s$ and $\eta(x) = t$, and denoted their gradients by $p = \nabla \xi$ and $q = \nabla \eta$, which (when linearly independent) span the 2D normal subspace associated with the two constraints in configuration space.
The only requirements that define $A$ and $B$, given in \cref{eq:codimension2_ibp_constraints_on_A,eq:codimension2_ibp_constraints_on_B}, are dot products with $p$ and $q$.
It is therefore a natural choice to take $A$ and $B$ to be inside $\mathrm{span}(\{p,q\})$ (in particular, neglecting tangent components to both constraints), meaning that 
\begin{equation}
    A = ap + bq \qquad \text{and} \qquad B = cp + dq,
\end{equation}
for $a,b,c,d \in \mathbb{R}$.
Substituting first this form of $A$ into the dot product constraints given in \cref{eq:codimension2_ibp_constraints_on_A} gives the following two equations
\begin{align}
    (ap + bq) \cdot p = a(p \cdot p) + b(q \cdot p) = 1, \\
    (ap + bq) \cdot q = a(p \cdot q) + b(q \cdot q) = 0, \\
\end{align}
which in matrix form reads
\begin{equation}
    G
    \begin{pmatrix}
        a \\ 
        b
    \end{pmatrix}
    =
    \begin{pmatrix}
        1 \\
        0
    \end{pmatrix}
    ,
    \quad 
    G = 
    \begin{pmatrix}
        p \cdot p & q \cdot p \\
        p \cdot q & q \cdot q
    \end{pmatrix}
    ,
\end{equation}
where $G$ is the Gram matrix.
Solving for the coefficients 
\begin{equation}
    \begin{pmatrix}
        a \\ 
        b
    \end{pmatrix}
    =
    G^{-1}
    \begin{pmatrix}
        1 \\
        0
    \end{pmatrix}
    ,
    \quad
    G^{-1} =
    \frac{1}{|G|} 
    \begin{pmatrix}
        q \cdot q & - q \cdot p \\
        - p \cdot q & p \cdot p
    \end{pmatrix}
    ,
\end{equation}
where $|G| = \mathrm{det}(G) = (p \cdot p)(q \cdot q) - (p \cdot q)^2$, we obtain
\begin{equation} \label{eq:codimension_2_A_field_def}
    \boxed{
    A = \frac{(q \cdot q) p - (p \cdot q) q}{|G|}
    }
    \ .
\end{equation}
Proceeding similarly for $B$, using \cref{eq:codimension2_ibp_constraints_on_B}, we obtain
\begin{equation} \label{eq:codimension_2_B_field_def}
    \boxed{
    B = \frac{-(q \cdot p) p + (p \cdot p) q}{|G|}
    }
    \ .
\end{equation}
Selecting $A$ to satisfy $A \cdot p = 1$ and $A \cdot q = 0$ makes $A$ have unit flux through the $\xi$-level set while remaining tangent to the $\eta$-level set, while selecting $B$ with $B \cdot p = 0$ and $B \cdot q = 1$ enforces the complementary property.
This is essential because a single application of the divergence theorem produces codimension-1 boundary terms, while the dual constraints imposed isolate the desired boundary (intersection) contribution. \\

\textbf{Higher codimension.}
The procedure obviously generalizes to observables with $k$-delta functions, of the form
\begin{equation}
    g(\mathbf{s}) = \left\langle \prod_{a=1}^k \delta(s_a - \xi_a(x)) \right\rangle,
\end{equation}
where the gradients $p_a = \nabla \xi_a \in \mathbb{R}^d$ can be defined.
However, eliminating delta functions sequentially increases the highest derivative order in the estimator linearly with $k$, which quickly becomes impractical as $k$ grows.
A compromise would be to eliminate one (or a few) delta functions analytically and treat the remaining variables with the usual binning or kernel smoothing approaches.

\vspace{0.2cm}
\begin{mybox}{Example: angular distribution functions}{codimension_2_example_adf}
A common structural observable is an angular distribution function (and its variants) built from particle triplet indices $(i,j,k)$.
One convenient codimension-2 example is the joint distribution of a bond length and an angle,
\begin{equation}
    C(s,t) \propto \left\langle \sum_{(i,j,k)}
    \delta(s - r_{ij}) \delta(t - c_{ijk})
    \right \rangle,
    \qquad
    c_{ijk} \equiv \cos\theta_{ijk} = \frac{\mathbf{r}_{ij} \cdot \mathbf{r}_{ik}}{r_{ij} r_{ik}}.
\end{equation}
Here $\xi = r_{ij}$ and $\eta = c_{ijk}$, so \eqref{eq:codimension_2_identity} converts the intersection estimator supported on $\{r_{ij} = s\} \cap \{c_{ijk} = t\}$ into a volume average over $\{r_{ij} \le s, \ c_{ijk} \le t\}$ plus drift/divergence terms, yielding an unbiased variance-reduced alternative to direct binning.
It is possible to derive the associated $A$ and $B$ vector fields using \cref{eq:codimension_2_A_field_def,eq:codimension_2_B_field_def} respectively.
\end{mybox}
\vspace{0.2cm}

\section{Periodic alternative for the $\mathbf{Q}$-function} \label{app:ac_pbc_alternative}

In this section, we present the estimator proposed by Qian \textit{et al.} \cite{2024_qian_space_warp_forces} that we compared with in the main manuscript, in the figure comparing various Hellmann-Feynman force estimators.
Instead of providing an explicit expression for the $\mathbf{Q}$-function (as we do in the main text), the authors implicitly define $\mathbf{Q}$ by providing an expression for its gradient.
The expression reads as follows 
\begin{equation}
\partial_{i \alpha} Q_{I\beta} = - \sum_\mathbf{T} Z_I \left[ \frac{\mathrm{erfc}(\kappa r_{iI\mathbf{T}})}{r_{iI\mathbf{T}}^3} r_{iI\mathbf{T},\beta} r_{iI\mathbf{T},\alpha} - \delta_{\alpha \beta} \frac{\mathrm{erfc}(\kappa r_{iI\mathbf{T}})}{r_{iI\mathbf{T}}} - \delta_{\alpha \beta} \frac{\kappa}{\sqrt{\pi}} \mathrm{Ei}(-\kappa^2 r_{iI\mathbf{T}}^2) \right],
\end{equation}
where $r_{iI\mathbf{T}} \equiv |\mathbf{r}_i - \mathbf{R}_I - \mathbf{T}|$, $\partial_{i \alpha} \equiv \partial / \partial r_{i\alpha}$, and $\alpha,\beta \in \{x,y,z\}$.
In vector form, the gradient of the $x$-component, for example, can be written explicitly as 
\begin{equation}
\nabla_i Q_{Ix} = - \sum_\mathbf{T} Z_I
\begin{pmatrix}
\displaystyle
\frac{\mathrm{erfc}(\kappa r_{iI\mathbf{T}}) r_{iI\mathbf{T},x}^2}{r_{iI\mathbf{T}}^3}
- \frac{\mathrm{erfc}(\kappa r_{iI\mathbf{T}})}{r_{iI\mathbf{T}}}
- \frac{\kappa}{\sqrt{\pi}} \mathrm{Ei}(-\kappa^2 r_{iI\mathbf{T}}^2)
\\[2.5ex]
\displaystyle
\frac{\mathrm{erfc}(\kappa r_{iI\mathbf{T}}) r_{iI\mathbf{T},x} r_{iI\mathbf{T},y}}{r_{iI\mathbf{T}}^3}
\\[2.5ex]
\displaystyle
\frac{\mathrm{erfc}(\kappa r_{iI\mathbf{T}}) r_{iI\mathbf{T},x} r_{iI\mathbf{T},z}}{r_{iI\mathbf{T}}^3}
\end{pmatrix}.
\end{equation}
The Laplacian of $Q_{I\beta}$ with respect to $r_i$ is obtained by taking the divergence of the above expression, i.e. $\nabla_i^2 Q_{I \beta} = \nabla_i \cdot \nabla_i Q_{I \beta}$.
Carrying out the differentiation analytically, one finds the closed-form expression
\begin{equation} \label{eq:bytedance_AC_laplacian}
    \nabla_i^2 Q_{I \beta} = -2 \sum_\mathbf{T} Z_I r_{iI\mathbf{T},\beta} \left[\frac{\mathrm{erfc}(\kappa r_{iI\mathbf{T}})}{r_{iI\mathbf{T}}^3} - \frac{\kappa}{\sqrt{\pi}}\frac{\exp(-\kappa^2 r_{iI\mathbf{T}}^2)}{r_{iI\mathbf{T}}^2} \right].
\end{equation}
While not directly provided in the aforementioned paper, the associated $\mathbf{Q}$-function components are given by
\begin{equation}
    Q_{I\beta}(\mathbf{r}) = Z_I \sum_i \sum_\mathbf{T} r_{iI\mathbf{T},\beta} \left( \frac{\mathrm{erfc}(\kappa r_{iI\mathbf{T}})}{r_{iI\mathbf{T}}} + \frac{\kappa}{\sqrt{\pi}} \mathrm{Ei}(-\kappa^2 r_{iI\mathbf{T}}^2) \right),
\end{equation}
where the exponential integral (Ei) term is an extra contribution that is absent in our definition of the $\mathbf{Q}$-function (which only involves the first term).

\end{document}